\documentclass[usenatbib]{mnras}
\usepackage{aas_macros}
\usepackage{amsmath}
\usepackage{amssymb}
\usepackage{color}
\usepackage{epsfig}
\usepackage{float}
\usepackage{graphicx}
\usepackage{latexsym}
\usepackage{morefloats}
\usepackage{natbib}
\usepackage{subfigure,multirow}
\usepackage{times}
\usepackage{ulem}

\newcommand{\plotone}[1]{\resizebox{0.95\hsize}{!}{\includegraphics{#1}}}

\newcommand{\plottwo}[2]{\resizebox{0.95\hsize}{!}
{\includegraphics{#1}\hspace{.5cm}\includegraphics{#2}}}

\newcommand{\plotx}[1]{\resizebox{0.75\hsize}{!}{\includegraphics{#1}}}

\newcommand{\Msun}{{~\rm M_\odot}}

\newcommand{\kpc}{~\rm kpc}
\newcommand{\Mpc}{~\rm Mpc}

\newcommand{\uvec}[1]{\boldsymbol{\mathit{\hat{#1}}}}

\def\gsim { \lower .75ex \hbox{$\sim$} \llap{\raise .27ex \hbox{$>$}}}
\def\lsim { \lower .75ex \hbox{$\sim$} \llap{\raise .27ex \hbox{$<$}}}
\newcommand{\eagle}{\textsc{eagle}}

\newcommand{\simRef}{Ref-L{\small 0100}N{\small 1504}}

\newcommand{\refeq}[1]{Eq. \eqref{#1}}
\newcommand{\refsec}[1]{Section \ref{#1}}
\newcommand{\reftab}[1]{Table \ref{#1}}
\newcommand{\reffig}[1]{Fig.~\ref{#1}}

\newcommand{\MWorbit}{\textit{MW-like-orbits}}

\definecolor{blue-violet}{rgb}{0.54, 0.17, 0.89}
\definecolor{purple}{rgb}{0.5, 0., 0.5}

\title[]
{The twisted dark matter halo of the Milky Way}
\author[Shao et al.]
{\parbox{\textwidth}{
 Shi Shao$^{1}$\thanks{E-mail: shi.shao@durham.ac.uk}, 
 Marius Cautun$^{2,1}$, 
 Alis Deason$^{1}$ and
 Carlos S. Frenk$^{1}$ \vspace{.20cm}}\\
$^1$Institute for Computational Cosmology, Department of Physics, Durham University, South Road Durham DH1 3LE, UK \\
$^2$ Leiden Observatory, Leiden University, PO Box 9513, NL-2300 RA Leiden, the Netherlands \\
}

\pubyear{2020}

\begin{document}
\label{firstpage}
\pagerange{\pageref{firstpage}--\pageref{lastpage}}
\maketitle

\begin{abstract}
    We analyse systems analogous to the Milky Way (MW) in the \eagle{} cosmological hydrodynamics simulation in order to deduce the likely structure of the MW's dark matter halo. We identify MW-mass haloes in the simulation whose satellite galaxies have similar kinematics and spatial distribution to those of the bright satellites of the MW, specifically systems in which the majority of the satellites (8 out of 11) have nearly co-planar orbits that are also perpendicular to the central stellar disc. We find that the normal to the common orbital plane of the co-planar satellites is well aligned with the minor axis of the host dark matter halo, with a median misalignment angle of only $17.3^\circ$. Based on this result, we infer that the minor axis of the Galactic dark matter halo points towards $(l,b)=(182^\circ,-2^\circ)$, with an angular uncertainty at the 68 and 95 percentile confidence levels of 22$^\circ$ and 43$^\circ$ respectively. Thus, the inferred minor axis of the MW halo lies in the plane of the stellar disc. The halo, however, is not homologous and its flattening and orientation vary with radius. The inner parts of the halo are rounder than the outer parts and well aligned with the stellar disc (that is the minor axis of the halo is perpendicular to the disc). Further out, the halo twists and the minor axis changes direction by $90^\circ$. This twist occurs over a very narrow radial range and reflects variations in the filamentary network along which mass was accreted into the MW.
\end{abstract}


\section{Introduction}
One of the fundamental predictions of the standard cosmological model
($\Lambda$CDM) is that galaxies are surrounded by extended
distributions of dark matter (DM) -- the DM haloes
\citep{Davis1985}. These are essential for galaxy formation since they
provide the gravitational potential wells within which gas is able to
cool, condense and form stars (\citealt{White1978,White1991}; for a
review see \citealt{Somerville2015}). DM haloes are the end product of
the anisotropic gravitational collapse of non-dissipative matter and
thus have highly non-spherical shapes
\citep[see][for recent reviews]{Frenk2012,Zavala2019}. Measuring the DM mass
distribution and, in particular, the shape of haloes, provides a
crucial test of the standard cosmological model and could reveal the
nature of DM or rule out alternative cosmological theories. Here, we
investigate how the Milky Way (MW) disc of satellite galaxies can be
used to infer the orientation and aspects of the formation history of
the Galactic DM halo.

Our galaxy offers a prime test-bed for characterising the DM
distribution around galaxies. Numerous studies have focused on
determining the mass and radial density profile of the Galactic DM
halo by analyzing the dynamics of halo stars, globular clusters and
satellite galaxies \citep[e.g.][]{Xue2008,Deason2012,Posti2018,
Callingham2019,Eadie2019,Watkins2019} or simply the number
and other properties of the satellites
\citep[e.g.][]{Busha2011,Cautun2014a}. By contrast, far fewer studies
have attempted to infer the shape and orientation of the Galactic DM
halo, which, in part, is a manifestation of the difficulties inherent
in such a task.

In $\Lambda$CDM, DM haloes have a range of shapes and can be described
as ellipsoidal mass distributions, with a preference for prolate over
oblate shapes \citep[e.g.][]{Frenk1988,Dubinski_1991,Warren_1992,
Jing_2002,Allgood_2006,Bett2007,Hayashi2007,Schneider2012}. The
axes ratios and orientations of the mass ellipsoids vary as a function
of distance from the halo centre and contain imprints of the past
growth history of the halo, with each shell retaining memory of the
mass accretion properties at the time when it collapsed
\citep[e.g.][]{Wechsler2002,Vera-Ciro2011,Wang2011a,Ludlow2013}. Galaxy
formation simulations have shown that the mass distribution within
haloes can be significantly affected by the baryonic distribution and,
in particular, by the orientation of the central galaxy. In the very
inner few tens of kiloparsecs, baryonic matter can dominate the
potential and cause the DM distribution to become less aspherical than
predicted by simulations of dissipationless collapse and well aligned
with the central galaxy
\citep[e.g.][]{Abadi2003,Bailin_2005,Bryan2013,Tenneti2014,Tenneti2015,
Velliscig2015a,Velliscig2015b,Shao2016,Chua2019}.
At large distances the potential of the baryonic component is
subdominant and the DM haloes retain a similar shape and orientation
to those found in DM-only simulations.

Since DM cannot yet be observed directly, the shape and orientation of
haloes can only be inferred from gravitational effects and
correlations with visible tracers. The wealth of dynamical tracers
around the MW and, in particular, the exquisite quality and sheer size
of the \textit{Gaia} dataset \citep{Gaia2018} has lead to the
development of a multitude of methods for studying the Galactic DM halo
\citep[see][for a recent review]{Wang2019}, including inferring halo
shapes from the properties of stellar streams
\citep[e.g.][]{Sanders2013,Price-Whelan2014,Bowden2015,Bovy2016,Malhan2019}, the
stellar halo \citep[e.g.][]{Bowden2016,Wegg2019} and hypervelocity
stars \citep[e.g.][]{Gnedin2005,Contigiani2019}.

Many studies of the shape of the Galactic DM halo are based on the
tidal stream of the Sagittarius dwarf, which traces the Galactic
potential within ${\sim}100\kpc$, and argue for a highly flattened
halo that is oriented perpendicular to the MW disc
\citep{Helmi2004,Johnston2005,Law2010,Deg2013}. The best fitting
\citet{Law2010} model has an oblate halo, with axes ratios,
$\langle c/a \rangle=0.72$ and $\langle b/a \rangle=0.99$, flatter
than the typical halo in $\Lambda$CDM \citep{Hayashi2007}; furthermore, its
alignment with the MW disc does not form a stable configuration
\citep{Debattista2013}. Motivated by these inconsistencies,
\citet{Vera-Ciro2013} improved the model by allowing the shape and
orientation of the DM halo to vary with radius, from a mildly
flattened halo in the inner ${\sim}20\kpc$ \citep[which is also
supported by GC dynamics,][]{Posti2018} to the \citeauthor{Law2010}
configuration at larger distances. \citeauthor{Vera-Ciro2013} and
\citet{Gomez2015} have highlighted that the Large Magellanic Cloud
(LMC), which is thought to be very massive
\citep{Penarrubia2016,Shao2018,Laporte2018,Cautun2019}, can induce significant
dynamical perturbations to the orbit of the Sagittarius tidal stream
as well as other streams \citep[e.g. the Tucana III
stream,][]{Erkal2018a}, thus further complicating the modelling of the
Galactic halo potential.

Most studies of halo shape and orientation are restricrted to the
inner DM halo ($<100\kpc$) since this is where the majority of
dynamical tracers are found. At larger distances little is known
about the shape of the halo and most conclusions are deduced from
statistical correlations. For instance, the central galaxy seems well
aligned with the inner halo and it has been argued that this alignment
is preserved, although with some degradation, all the way to the
virial radius, allowing the orientation of the halo minor axis to be
inferred within a median angle of ${\sim}33^\circ$
\citep[e.g.][]{Bailin_2005,Tenneti2015,Velliscig2015a,Shao2016}.

Satellite galaxies are preferentially accreted along filaments
\citep{Libeskind2005,Libeskind2014,Shao2018a} -- in the same
directions as mass is accreted onto haloes -- and thus the satellites
also trace the DM halo including its large-scale orientation
\citep{Libeskind2007,Shao2016}. However, in the MW the satellites are
found in a plane perpendicular to the Galactic disc
\citep[e.g.][]{Kunkel1976, Lynden-Bell1976, Lynden-Bell1982,
Kroupa2005}, and this suggests a very different halo orientation
from that inferred from the orientation of the MW
disc. \citet{Shao2016} studied configurations in which the
satellites are found in a plane perpendicular to the central disc and
have found that, in this case, the DM halo is poorly aligned with the
central galaxy. Thus, we cannot use the MW stellar disc to predict the
orientation of the Galactic halo.

In this paper we use the rotating disc of classical dwarf galaxies in
the MW to infer possible formation histories and configurations of the
Galactic DM halo. The paper is motivated by the results of
\citet{Shao2019} who showed that, out of the 11 MW classical dwarfs, 8
orbit in nearly the same plane \citep[see also][]{Pawlowski2013} --
specifically the orbital poles of those 8 satellites are enclosed
within a $22^\circ$ opening angle. \citet{Shao2019} showed that
MW-like rotating planes of satellites in $\Lambda$CDM are a
consequence of highly anisotropic accretion and, most importantly for
this study, of the torques exerted by the host halo which tilt the
satellite orbits onto the host halo's equatorial plane. This suggests
that the satellite orbital plane should be a good indicator of halo
orientation, which is one of the main questions we investigate here.

We proceed by identifying in the \eagle{} galaxy formation simulation
\citep{Schaye2015} satellite systems similar to the MW, in which 8 out
of the brightest 11 satellites orbit in nearly the same plane. The
common orbital plane is very nearly perpendicular to the minor axis of
the host DM halo and thus can be used to predict the orientation of
the Galactic DM halo. We then identify \eagle{} MW-mass systems which
have a rotating plane of satellites that is perpendicular to their
central galaxy, as found in our Galaxy, and perform an in-depth study
of such systems. The goal is to understand the processes that give
rise to the perpendicular configuration between satellites and central
galaxy and what this can tell us about the formation history of the
Galactic DM halo.

The paper is organised as follows. In Section~\ref{sect:simul} we
review the simulations used in this work and describe our sample
selection; in Section~\ref{sect:result} we analyse the DM halo
properties of systems which have satellite distributions similar to
our own galaxy; then in Section~\ref{sect:MW-like_examples} we study the formation
history of five MW-mass haloes that are very similar to the MW; we
conclude with a short summary and discussion in
Section~\ref{sect:conclusions}.

\section{Simulation and sample selection}
\label{sect:simul}

We analyze the main cosmological hydrodynamics simulation (labelled
\simRef{}) of the \eagle{} project \citep{Schaye2015,
Crain2015}. The simulation follows the evolution of a periodic cube of
sidelength $100\Mpc{}$ with $1504^3$ DM particles and an initially
equal number of gas particles. The DM particle mass is
$9.7\times 10^6 \Msun$ and the initial gas particle mass 
$1.8\times 10^6 \Msun$. The simulation assumes the 
\textit{Planck} cosmological parameters \citep{Planck2014}:
$\Omega_{\rm m}=0.307, \Omega_{\rm
 b}=0.04825,\Omega_\Lambda=0.693,h=0.6777,\sigma_8=0.8288$ and
$n_{\rm s}=0.9611$.

The simulation was performed with a modified version of the
\textsc{gadget} code \citep{Springel2005}, which includes
state-of-the-art smooth particle hydrodynamics and subgrid models for
baryonic processes such as element-by-element gas cooling, star
formation, metal production, stellar winds, and stellar and black hole
feedback. The \eagle{} subgrid models were calibrated
to reproduce three present-day observables: the stellar mass function,
the dependence of galaxy sizes on stellar mass, and the normalization
of the relation between supermassive black hole mass and host galaxy
mass. For a more detailed description please see \citet{Schaye2015}
and \citet{Crain2015}.

To identify analogues of the MW satellite system we make use of
the $z{=}0$ \eagle{} halo and galaxy catalogue
\citep{McAlpine2016}. The haloes and galaxies correspond to
gravitationally bound substructures identified by the
\textsc{subfind} code \citep{Springel2001,Dolag2009} applied to the
full mass distribution (DM, gas and stars). The main haloes are
characterized by the mass, $M_{200}$, and radius, $R_{200}$, that
define an enclosed spherical overdensity of $200$ times the critical
density. The position of each galaxy, both centrals and
satellites, is given by their most bound particle. We also study the
formation history of several individual systems, for which we use the
\eagle{} galaxy merger trees \citep{Qu2017} built from over 200
snapshots (roughly one every 70 Myrs).

\subsection{Sample selection}

We wish to work with a cosmologically representative sample of
satellite systems of similar stellar masses to the classical dwarfs
that orbit around the MW. We use the sample studied by
\citet{Shao2016} which consists of 1080 \eagle{} galaxies selected
according to the following criteria: i) they are the central
galaxy in a halo of mass $M_{200}\in[0.3,3]\times10^{12}\Msun$ and
ii) they have at least 11 luminous satellites within a
distance of $300 \kpc$ from the central galaxy. We define luminous
satellites as subhaloes with at least one associated stellar
particle, corresponding to objects of stellar mass larger than
${\sim}1\times10^6\Msun$. If there are more than 11 satellites
within the chosen distance, we only consider the 11 with the highest
stellar mass. 

We further require that the MW-like analogues be isolated, that is
that they have no neighbours more massive than themselves within a
distance of $600\kpc$. This isolation criterion is not
very strict and, for example, the MW would fulfill it since
Andromeda is ${\sim}800 \kpc$ away \citep{McConnachie2012}. The
median halo mass of our sample is $1.2\times10^{12}\Msun$ (see
Fig.~A1 of \citealt{Shao2016} for the exact halo mass distribution),
which is in good agreement with recent determinations of the MW halo
mass (e.g. see \citealt{Patel2018,
Deason2019,Callingham2019,Cautun2020}; and Fig.~5 in the
\citealt{Wang2019} review). We have not applied a
morphological selection, so our sample of MW-like galaxies
contains the full range of morphologies produced in the \eagle{}
model \citep{Trayford2015,Correa2017}.

The radial distribution of satellites in our \eagle{}
MW-mass sample is slightly less concentrated than the distribution
of the MW classical satellites \citep[see
also][]{Yniguez2014,Carlsten2020}. This is likely due to
substructures that are prematurely disrupted in the simulation
\citep{Bose2020}. These absent satellites should typically be
found in the inner $100\kpc$ of the halo and, comparing to the
radial distribution of \citeauthor{Bose2020}, we estimate that our
sample is lacking, on average, 1-2 such satellites per system. We
expect that including these additional satellites in future
studies based on much higher resolution simulations would have
minimal impact on our conclusions. Most satellites are accreted
along a plane \citep{Shao2019}, so the inner satellites have, on
average, the same distribution of orbital planes as the outer
ones. Furthermore, as we shall discuss shortly, our study concerns
planes defined by 8 out of the 11 brightest satellites that
have the most co-planar orbits. This definition results in planes
that are robust to replacing one of the satellites since, even if
the newly added object had a very different orbital plane, it
would not be included in our subset of satellites with co-planar
orbits and thus would not have a large impact on the plane of
satellite galaxies.


\begin{figure*}
    \plotx{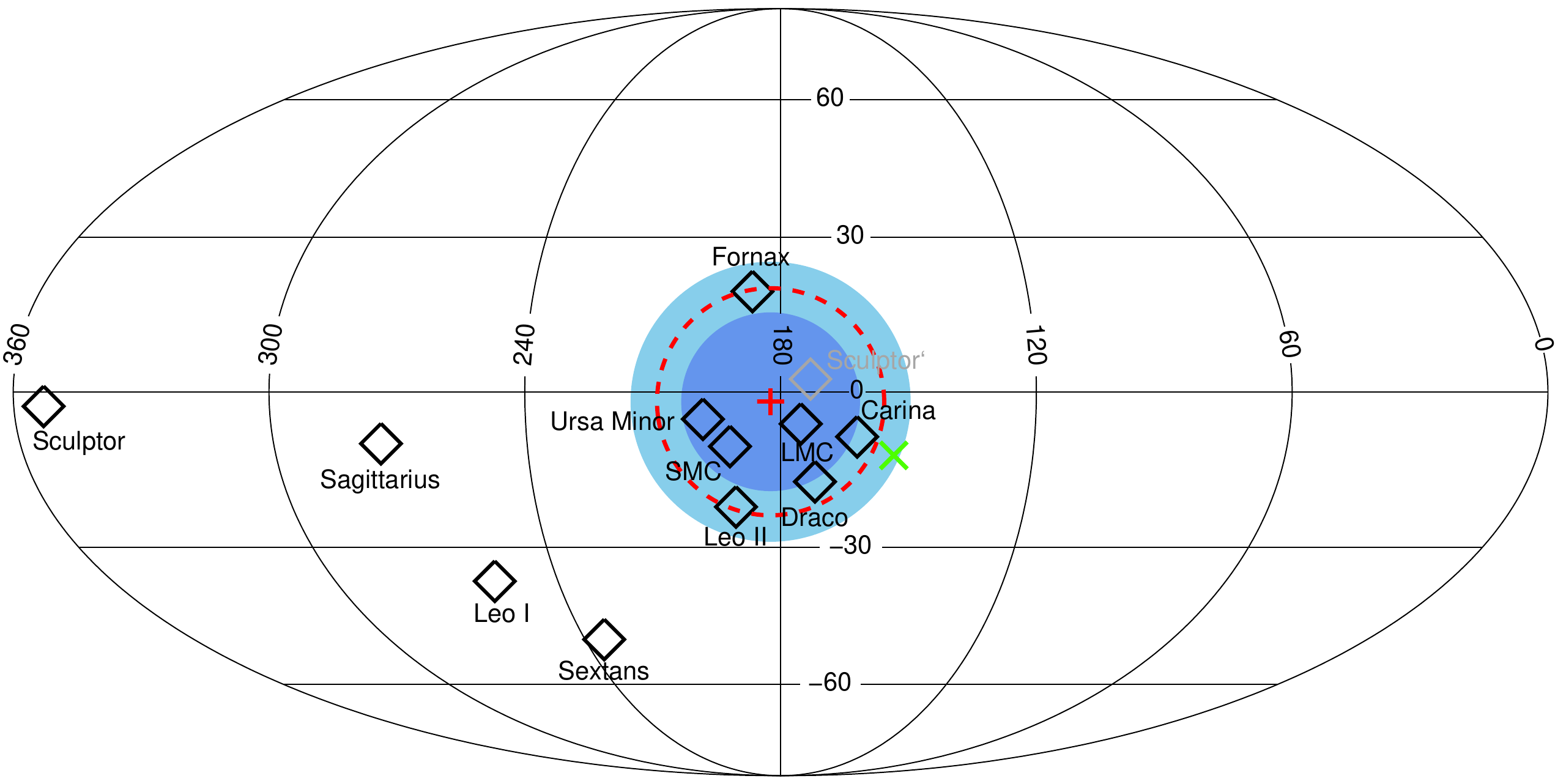}
	\caption{Aitoff projection of the orbital poles of the 
          classical satellites of the MW. Each black rhombus corresponds to the
          orbital pole of a classical dwarf in Galactic longitude,
          $l$, and latitude, $b$. \citet{Shao2019} have shown that 8
          out of the 11 classical satellites have highly clustered
          orbital poles that are contained within a $22^\circ$ opening
          angle (red dashed circle) around the direction
          $(l,b)=(182^\circ,-2^\circ)$ (red cross symbol). Out of the
          8 satellites with co-planar orbits, Sculptor is
          counter-rotating, and, to emphasise that its orbit is in the
          same plane, the grey rhombus shows its position after
          flipping its orbital pole. The green x shows the
          minor axis of the spatial distribution of
          satellites. \textit{In this paper, we 
          show that the minor axis of the Galactic DM halo likely
          points towards the red cross symbol.} The two coloured
          regions show respectively the 50 and 75 percentile
          confidence interval for our determination of the orientation
          of the halo's minor axis. 
	}
	\label{fig:MW_sky}
\end{figure*}

\subsection{Identifying MW-like rotating planes of satellites}
\label{sec:methods}

We identify satellites with co-planar orbits using the method
introduced by \citet{Shao2019}. The goal is to find satellite
distributions similar to that in the MW, where 8 out of the 11
classical satellites orbit in roughly the same plane\footnote{The
 choice of 8 out of 11 satellites is explained in Fig.~3 of
 \citeauthor{Shao2019}, which shows that a subset of 8 classical MW
 dwarfs have highly co-planar orbits that stand out when compared to
 either typical $\Lambda$CDM systems or isotropic distributions of
 orbits.}. This is illustrated in \reffig{fig:MW_sky}, which shows
the orbital poles of the classical satellites. \cite{Shao2019} have
quantified the degree of coplanarity of the orbits by the minimum
opening angle, $\alpha_8$, needed to enclose the orbital poles of the 8
satellites whose orbits are closest to a single plane. For the MW,
$\alpha_8=22^\circ$, shown in \reffig{fig:MW_sky} as the red dashed
circle centred on $(l,b)=(182^\circ,-2^\circ)$.

For each MW-mass galaxy in our sample we identify the subset of 8
satellites whose orbits exhibit the highest degree of coplanarity as
follows. We first generate $10^4$ uniformly distributed directions on
the unit sphere and, for each direction, we find the minimum opening
angle that includes the orbital poles of 8 satellites. We then
select the direction with the smallest opening angle. We denote the
smallest opening angle as $\alpha_8$; its corresponding direction is
the normal to the \textit{common orbital plane} in which the 8
satellites orbit, which we denote as $\uvec{n}_{\rm{orbit}}$.

The distribution of $\alpha_8$ opening angles for $\Lambda$CDM MW-mass
haloes can be found in Fig~4 of \citet{Shao2019}.
We emphasise
that very few (only 6 out of 1080) \eagle{} haloes have $\alpha_8$
values as low as the MW. The rarity of such MW-like systems is
somewhat by construction, because we want to study a feature of the MW
satellite distribution that is uncommon when compared to the typical
$\Lambda$CDM halo. In fact, a considerable fraction of $\Lambda$CDM
haloes have rotating planes of satellites; however each plane is
different suggesting that the planes encode information about the
evolution of that particular system \citep{Cautun2015}. To obtain a
reasonable number of \eagle{} satellite systems with orbits similar to
those of the MW system, we define \MWorbit{} systems as those with
opening angles, $\alpha_8<35^\circ$. There are ${\sim}140$ \eagle{}
haloes (13\% of the sample) that fulfil this selection criterion.

\section{The DM haloes of \MWorbit{} systems}
\label{sect:result}
We refer to the DM haloes of galactic-mass systems in which 8 out
of the brightest 11 satellite galaxies orbit in a narrow plane as
\MWorbit{} systems We study the shape of the DM haloes and their
orientation relative to the plane of satellite, the central galaxy and
the large-scale structure surrounding these systems.

\subsection{The DM halo shape}
\label{subsec:DM_halo_shape}

\begin{figure*}
    \centering
	\plottwo{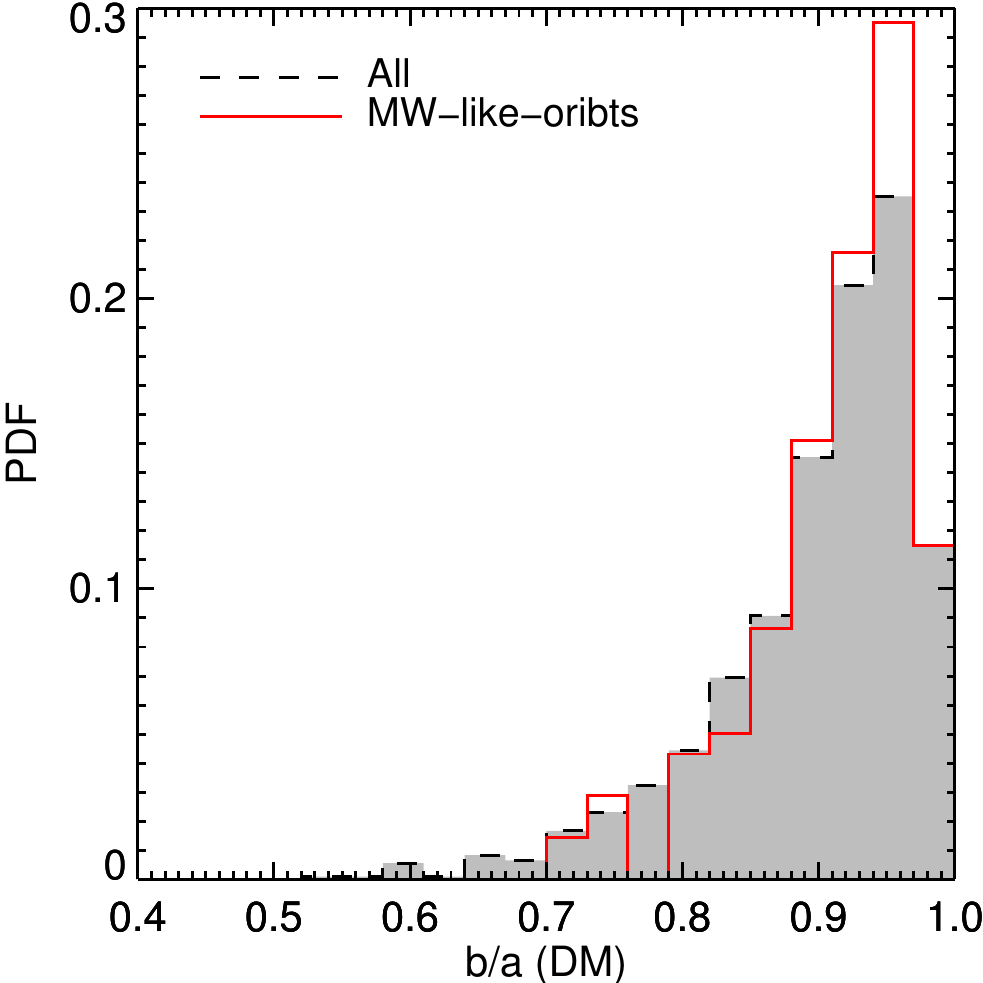}{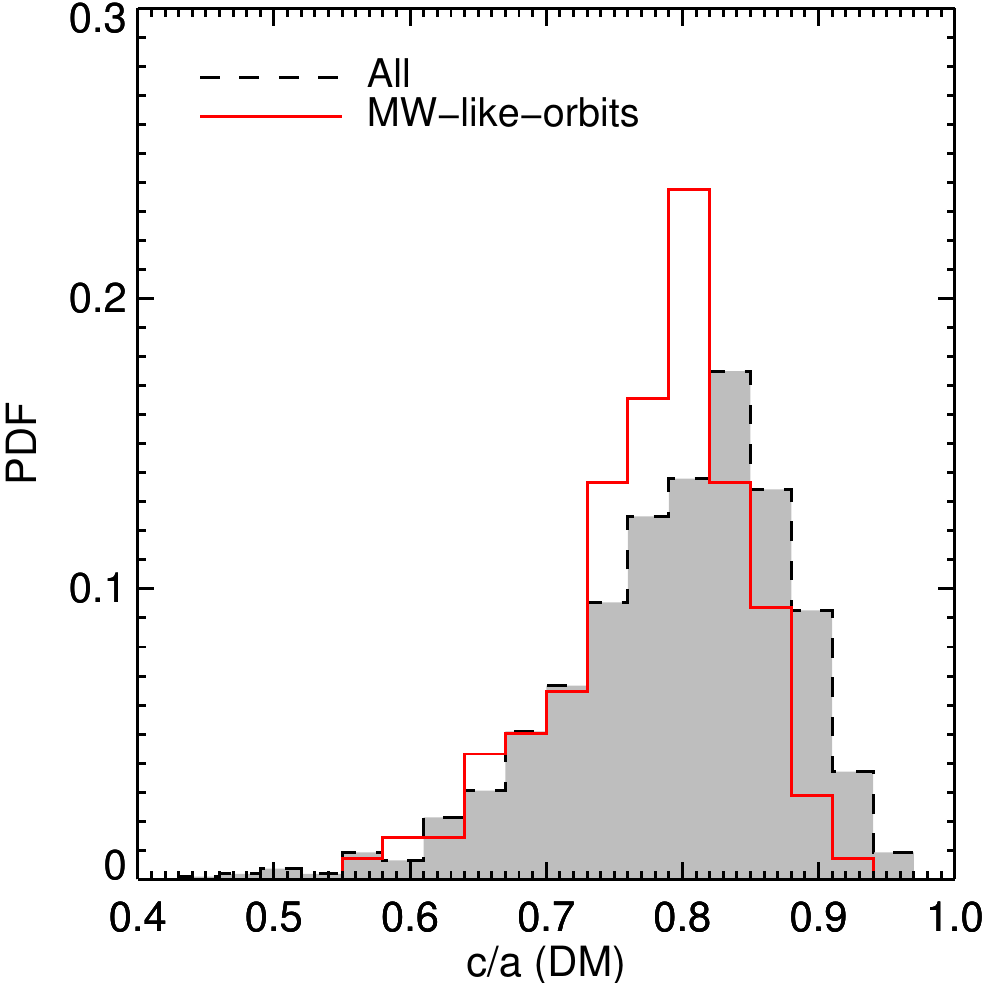}
	\caption{ The distribution of axes ratios, $b/a$ (left panel)
        and $c/a$ (right panel) of the shape of the entire DM halo,
        that is all the DM particles within $R_{\rm 200}$. The
        dashed line shows the result for all MW-mass haloes, while
        the red solid line corresponds to the sample of \MWorbit{}
        systems. We find with a high statistical confidence that
        the Galactic DM halo is more flattened (smaller $c/a$) than
        the average $\Lambda$CDM halo. Also, we find hints that the
        MW halo is more likely to have $a{\approx}b$ (i.e. more oblate)
        than the typical expectation, however due to the small sample
        size we cannot rule out that this difference is due to statistical
        fluctuations (see main text for details).
	}
	\label{fig:Pdf_ca_halo}
\end{figure*}

We characterise the shape of a DM halo by its mass tensor,
\begin{equation}
    I_{ij} \equiv \sum_{k=1}^{N} x_{k,i}\;x_{k,j}
    \label{eq:tensor} \;,
\end{equation}
where the sum is over the DM particles found within the halo radius,
$R_{200}$. The quantity $x_{k,i}$ denotes the $i$-th component
($i=1,2,3$) of the position vector associated with the $k$-th DM
particle, measured with respect to the halo centre. The
shape and the orientation are determined by the eigenvalues,
$\lambda_i$ ($\lambda_{1}\geqslant\lambda_{2}\geqslant\lambda_{3}$),
and the eigenvectors, $\uvec{e}_i$, of the mass tensor. The major,
intermediate and minor axes of the corresponding ellipsoid are given
by $a=\sqrt{\lambda_1}$, $b=\sqrt{\lambda_2}$, and
$c=\sqrt{\lambda_3}$, respectively, and their orientation is given by
$\uvec{e}_1$, $\uvec{e}_2$ and $\uvec{e}_3$.
We obtain the same eigenvalues and eigenvectors if, instead, we
define the halo shape using the moment of inertia tensor
\citep{Bett2007}.

When calculating halo shapes we use all the DM particles
enclosed within $R_{\rm 200}$ (for the full halo), or within a fixed
radial distance when calculating the shape as a function of radius.
We prefer this choice compared to alternatives such as removing
bound substructures, since it is closer to what can be done in
observations. Observations measure the shape of the total
gravitational potential and, with a few exceptions such as the LMC,
it is very difficult to isolate the contribution of each
substructure to the total potential. In general, the shape
measurement is mostly insensitive to substructures, except for a
small number of hosts that contain massive satellites
\citep{Bett2007}.

We describe the halo shape
by the intermediate-to-major, $b/a$, and minor-to-major,
$c/a$, axes ratios, which characterize the degree of halo flattening.
The two axes ratios are shown in \reffig{fig:Pdf_ca_halo}, where we
compare the flattening of the full sample of MW-mass systems to that
of \MWorbit{} ones. The full sample is characterised by preferentially
prolate haloes ($a>b \approx c$) with a median flattening of
$b/a{\sim}0.9$ and $c/a{\sim}0.8$, in good agreement with previous
studies \citep[e.g.][]{Frenk1988,Bett2007,Schneider2012,Shao2016}.

The haloes of the \MWorbit{} sample show systematic differences compared
to the full sample. Since the \MWorbit{} sample is rather small,
only 139 objects, we assess the statistical significance of any
observed differences using the Kolmogorov-Smirnov (KS) test. Firstly,
the \MWorbit{} haloes have $b/a$ axes ratios that are somewhat larger
than that of the full sample. However, the effect is rather small and
a KS-test indicates that the difference is not statistically significant
(i.e. there is a $p=0.27$ probability that both samples follow the same
distribution). Secondly, the \MWorbit{} haloes have $c/a$ axes ratios
that are systematically smaller than that of the full sample. This result
is statistically robust, with a KS-test probability of $p=3\times10^{-4}$
that the observed difference is due to statistical fluctuations. Thus,
the orbital clustering of the MW classical satellites indicates that the
Galactic DM halo is systematically flatter (i.e. smaller $c/a$ ratio) than
the typical $\Lambda$CDM halo.

The haloes of \MWorbit{} systems are flattened because they experience
a higher degree of anisotropic accretion, especially planar infall,
than the average $\Lambda$CDM halo. This is illustrated in Figure~7 of
\citet{Shao2019}, where we showed that systems with many co-planar
satellite orbits had a higher degree of anisotropic infall \citep[see
also][]{Libeskind2005,Lovell2011,Deason2011,Kang2015,Shao2018a}. The
preferential infall plane is responsible for the coherent orbital
planes of satellites as well as for the flattening of the DM halo,
with the equatorial plane of the halo being aligned with the
anisotropic infall plane
\citep[e.g.][]{Hahn2007,GaneshaiahVeena2018,GaneshaiahVeena2019}.

\subsection{The orientation of the DM halo with respect to the satellite distribution}
\label{subsec:results:sat-halo_alignment}
\begin{figure*}
	\plottwo{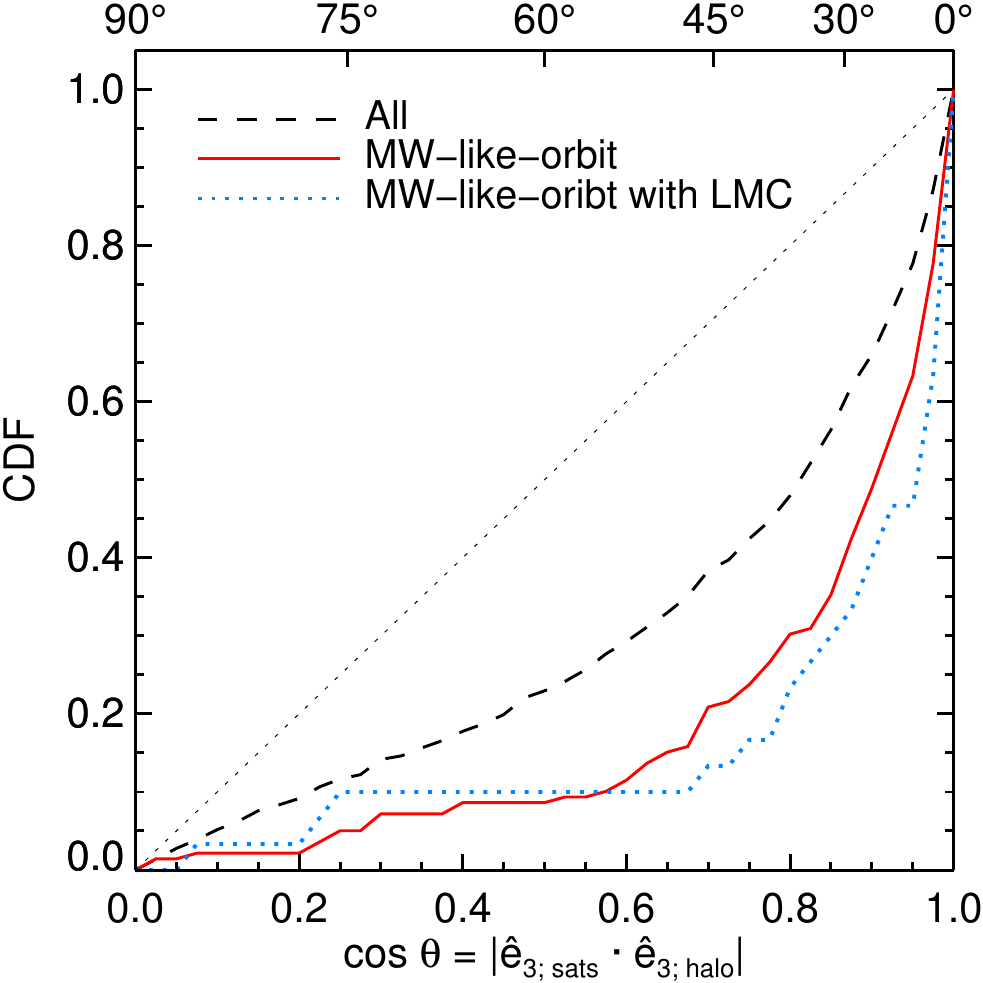}{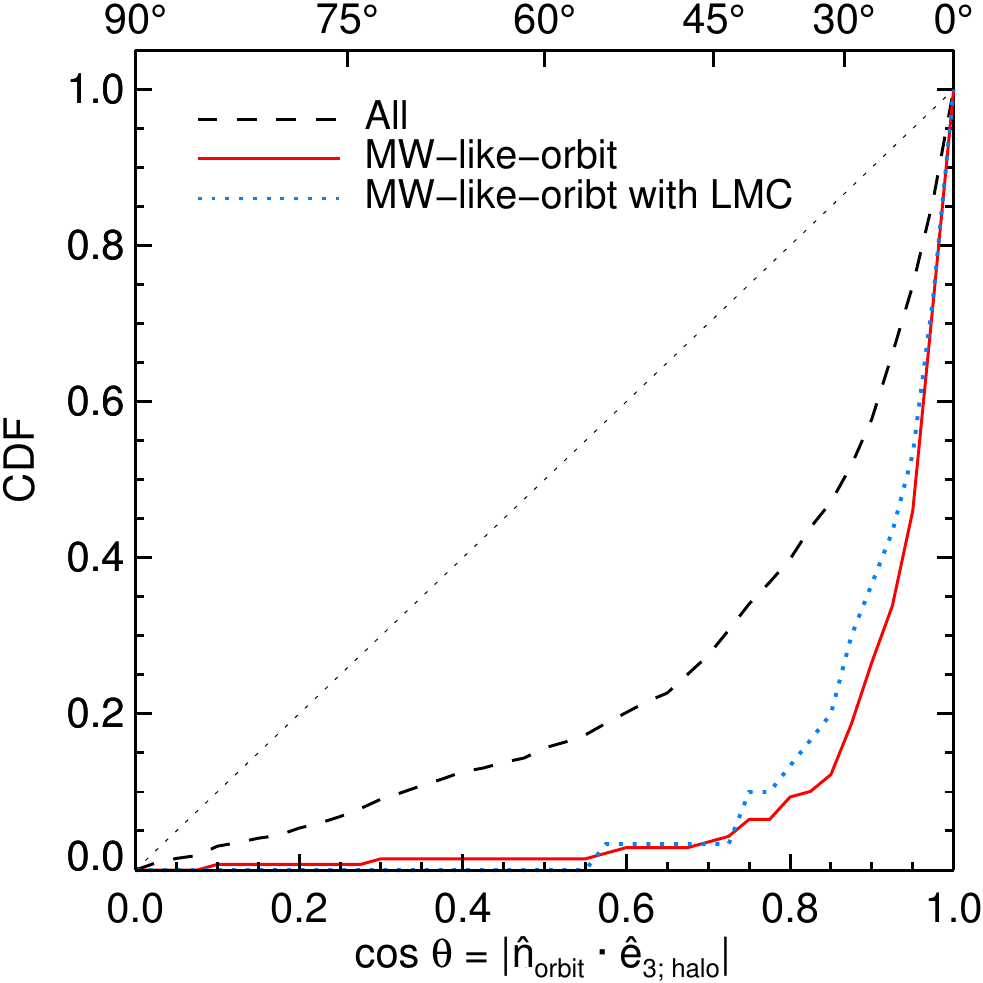}
	\caption{ \textit{Left panel:} the CDF of the alignment angle,
	$\cos\theta$, between the minor axis of the satellite system
    and the minor axis of the DM halo in the \eagle{} simulation.
    The three lines correspond to: all MW-mass haloes (dashed),
    \MWorbit{} haloes (solid red) and \MWorbit{} haloes with an
    LMC-mass dwarf satellite (dotted blue). \textit{Right panel:}
    as the left panel, but for the alignment angle between the normal
    to the common orbital plane of the satellites and the minor axis
    of the DM halo. In both panels, the dotted diagonal line shows
    the CDF for the no-alignment case. {Both the minor axis of the
    satellite system and the normal to the common preferential orbital
    plane are aligned with the halo minor axis; however, the latter
    shows a much tighter alignment. Thus, the plane in which most
    satellites orbit is a very good indicator of the DM halo minor
    axis and especially for systems that have \MWorbit{} planes.}
	}
	\label{fig:Sat_halo}
\end{figure*}

\begin{table}
    \centering
    \caption{ The median alignment angles between the satellite
      distribution of MW-mass galaxies and their DM halos, central
      galaxies, and surrounding large scale structure. We provide
      results for three samples of MW-mass systems: all (second
      column), those with \MWorbit{} (third column) and those with
      \MWorbit{} that also have an LMC-mass dwarf satellite (fourth
      column). We provide values for the median angle and the 68
      percentile confidence interval with which we can determine the median.
    }
   
    \renewcommand{\arraystretch}{1.3} 
    \begin{tabular}{  @{} p{.23\columnwidth} p{.14\columnwidth} p{.2\columnwidth} p{.27\columnwidth} @{}   } 
        \hline\hline
        Alignment type & \multicolumn{3}{c}{Sample} \\[-.1cm]
        \hline
         & All & \MWorbit{} & \MWorbit{} with LMC-mass satellite \\
         \hline
        (1) \ $\theta_{\rm sats - halo}$ & $35.7^{+0.9}_{-0.8}$ & $24.8^{+2.9}_{-1.4}$ & $16.2^{+8.1}_{-2.9}$ \\
        (2) \ $\theta_{\rm orbit - halo}$ & $30.2^{+0.7}_{-1.1}$ & $17.3^{+1.1}_{-0.6}$ & $18.7^{+3.9}_{-3.9}$ \\
        (3) \ $\theta_{\rm orbit - cen}$ & $41.3^{+1.5}_{-0.7}$ & $24.8^{+1.4}_{-0.9}$ & $36.2^{+7.5}_{-10.1}$ \\
        (4) \ $\theta_{\rm orbit - LSS}$ & $51.6^{+1.5}_{-1.6}$ & $45.1^{+2.0}_{-4.5}$ & $37.9^{+7.2}_{-2.6}$ \\
        
        \hline\hline
    \end{tabular}
    
    \vskip .1cm
    \renewcommand{\arraystretch}{1.3}
     \begin{tabular}{ @{} p{1\columnwidth} @{} }
        (1) - the median angle between the minor axis of the satellite
       distribution, $\uvec{e}_{\rm 3; \ sats}$, and the minor axis of
       the DM halo, $\uvec{e}_{\rm 3; \ halo}$. \\ 
        (2) - the median angle between the normal to the common
       orbital plane of the satellites, $\uvec{n}_{\rm orbit}$, and
       the minor axis of the DM halo, $\uvec{e}_{\rm 3; \ halo}$. \\ 
        (3) - the median angle between the normal to the common
       orbital plane of the satellites, $\uvec{n}_{\rm orbit}$, and
       the minor axis of the central galaxy, $\uvec{e}_{\rm 3; \
       cen}$. \\ 
        (4) - the median angle between the normal to the common
       orbital plane of the satellites, $\uvec{n}_{\rm orbit}$, and
       the first LSS collapse axis (i.e. the perpendicular to the LSS
       sheet), $\uvec{e}_{\rm LSS}$. \\ 
    \end{tabular}
    \renewcommand{\arraystretch}{1.0}
    \label{tab:median_alignment_angle}
\end{table}

We now study the extent to which the satellite distribution can constrain
the orientation of the DM halo. To this aim, \reffig{fig:Sat_halo}, shows
the alignment of the satellite distribution with the minor axis of the DM halo.

To begin with, we follow the standard approach in the literature and
define the orientation as the direction of the minor axis of the
satellite system
\citep[e.g.][]{Kroupa2005,Libeskind2005,Deason2011}. This is
calculated from the mass tensor of the distribution using
\refeq{eq:tensor} applied to the 11 brightest satellites of each
system. The resulting orientation of the minor axis of the MW
classical satellites is shown as the green cross symbol in
\reffig{fig:MW_sky}. Applying the same procedure to the \eagle{}
systems, we find a moderate alignment between the minor axis of the
satellite distribution and the minor axis of the DM halo, with a
median alignment angle of 35.7$^\circ$. The
subset of systems with \MWorbit{} show a better alignment between
their satellite distribution and DM halo, with a median alignment
angle of 24.8$^\circ$ (see \reffig{fig:Sat_halo}
and \reftab{tab:median_alignment_angle}).

The MW has recently accreted a massive satellite, the LMC, that could
potentially affect the orientation of its DM halo
\citep[e.g.][]{Garavito-Camargo2019} and satellite orbits
\citep[e.g.][]{Gomez2015,Patel2020}, in addition to bringing in its
own satellites. To study the potential effect of the LMC, we have
further identified the subset of \MWorbit{} system that also have an
LMC-mass dwarf satellite. We define an LMC-mass analogue as any
satellite located less than 150$\kpc$ from the central galaxy and with
stellar mass greater than $1\times10^9\Msun$
\citep{van_der_Marel2002,McConnachie2012,Shao2018}. We find that
${\sim}20\%$ of the sample (30 out of 139) have an LMC-mass dwarf;
we find roughly the same prevalence of LMC-mass satellites for the
full population of MW-mass systems. The alignment of \MWorbit{} systems
that have an LMC-mass satellite is similar to that of the \MWorbit{} subset,
with differences consistent with stochastic effects due to the small
number of systems with LMC-mass satellites.

The orientation of the satellite distribution can also be defined as
the normal, $\uvec{n}_{\rm orbit}$, to the common orbital plane of the
8 satellites with the most co-planar orbits (see
\refsec{sec:methods}). The $\uvec{n}_{\rm orbit}$ direction is robust,
varying only slowly with time. This is in contrast to the minor axis
of the satellite distribution, which can vary rapidly with time and
whose orientation is especially sensitive to the farthest most
satellites \citep[e.g.][]{Buck2016,Lipnicky2017,Shao2019}. The
alignment of $\uvec{n}_{\rm orbit}$ with the halo minor axis,
$\uvec{e}_{\rm 3; \ halo}$, is shown in the right-hand panel of
\reffig{fig:Sat_halo}. We find that, on average, $\uvec{n}_{\rm orbit}$
is better aligned with $\uvec{e}_{\rm 3; \ halo}$ than the satellites'
minor axis. This is true for both the full population of MW-mass
halos, and even more so for the \MWorbit{} systems, which have a
median angle between $\uvec{n}_{\rm orbit}$ and
$\uvec{e}_{\rm 3; \ halo}$ of only 17.3$^\circ$.

The very strong alignment between the normal to the common orbital
plane of satellites and the halo minor axis for \MWorbit{} systems
means that we can predict the orientation of the Galactic DM halo with
rather small uncertainty. The most likely orientation of the MW halo,
$\uvec{e}_{\rm 3; \ halo \; MW}$, corresponds to
$(l,b)=(182^\circ,-2^\circ)$ and the 50, 75 and 90 percentile
confidence intervals correspond to angles of 17.3$^\circ$, 26.9$^\circ$, and
36.6$^\circ$, respectively. This prediction is
shown in \reffig{fig:MW_sky} by the red cross symbol for the most
likely orientation of the halo minor axis, and by the two shaded
regions for the 50 and 75 percentile confidence intervals.

The MW has a close neighbour, M31, which is currently located 
${\sim}800\kpc$ away and is thought to be on first approach
\citep{vanderMarel2012}. The large distance between the two
galaxies, ${\sim}4$ times the MW halo radius,
$R_{200}$ \citep{Cautun2020}, and the fact that M31 is on first
approach make it very unlikely that the MW halo shape has been
affected directly by the presence of M31. However, as we will
discuss in \refsec{subsec:results:sat-LSS_alignment}, since  both
the MW and M31 form in the same large-scale tidal field, we expect
that the MW has a weak tendency to align with this tidal field and,
in turn, to align with the MW-M31 direction, which is, to a much larger
degree, determined by the large-scale tidal field (van Leeuwen in
prep.). We note that the alignment between halo shapes and the tidal
field is taken into account in \reffig{fig:Sat_halo} and, in fact,
it is one of the key processes responsible for the results shown in
the figure.

The alignment of the normal to the common orbital plane of satellites with the halo
minor axis is better than the galaxy--halo minor axis
alignment, which has a median angle of $33^\circ$
\citep[e.g.][]{Shao2016}, and thus provides a more robust way to infer
the DM halo orientation. This is especially the case for systems in
which many satellites have co-planar orbits, such as our Galaxy.

We note that the MW is an extreme object in the \MWorbit{} sample,
which has been selected to have opening angles $\alpha_8 < 35^\circ$,
while the MW has $\alpha_8^{MW} = 22^\circ$. We find that satellite
systems with $\alpha_8$ values similar to that of the MW (such systems
are very rare, with only 10 out of 1080 having $\alpha_8<25^\circ$)
show an even tighter alignment between the common orbital plane and
the DM halo orientation and thus potential future studies that have
access to larger cosmological simulations could constrain the Galactic
halo orientation even better. Here, we do not quote any numbers because
the small sample size precludes us obtaining statistically robust results.

\subsection{The alignment of satellite systems with their central galaxies}
\label{subsec:results:sat-cen_alignment}
\begin{figure}
	\plotone{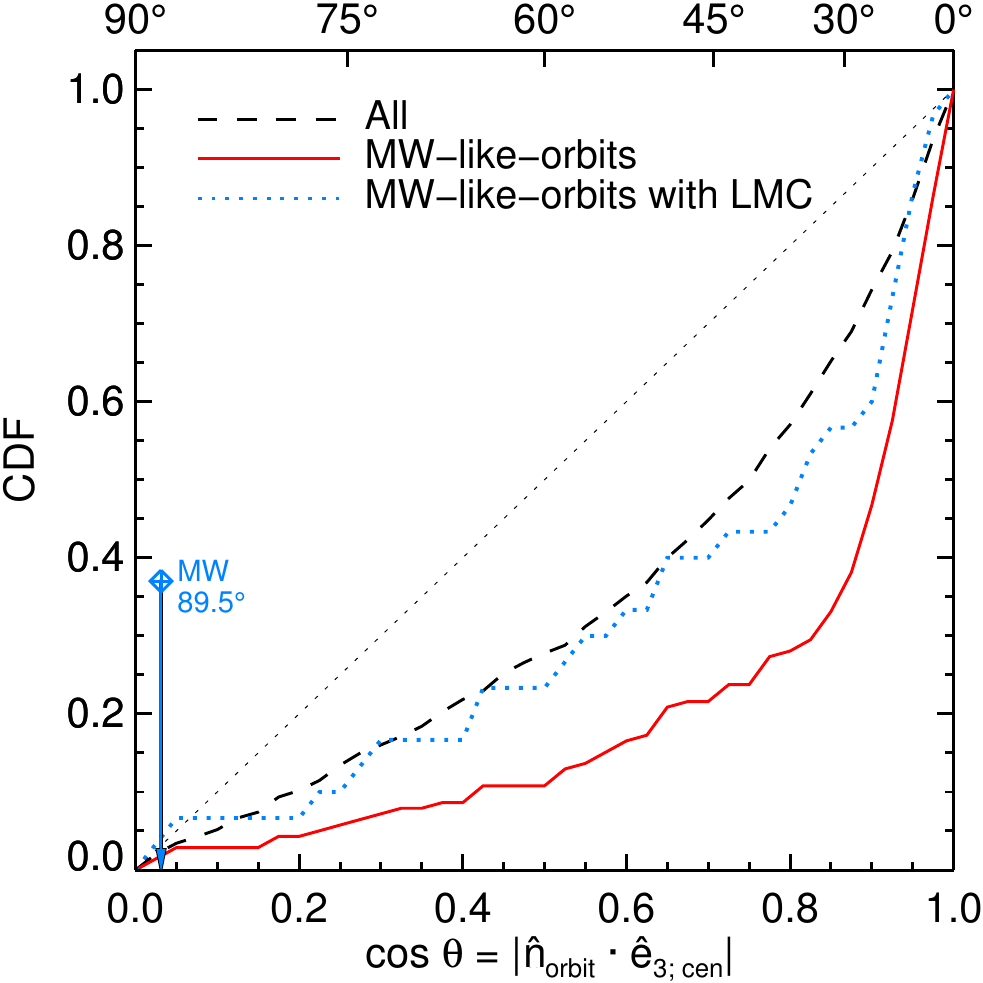}
	\caption{ The CDF of the alignment angle, $\cos \theta$,
          between the normal, $\uvec{n}_{\rm orbit}$, to the
          preferential orbital plane and the minor axis,
          $\uvec{e}_{\rm cen}$, of the stellar disc. The vertical
          arrow at $\cos\theta\approx0$ indicates the measured value
          for our galaxy. }
	\label{fig:Theta_a8_cen}
\end{figure}

In \reffig{fig:Theta_a8_cen} we study how the satellite systems are
oriented relative to the disc of the central galaxy. This is motivated
by our own Galaxy, where the common orbital plane of satellites is
perpendicular to the MW stellar disc (see \reffig{fig:MW_sky} where
the MW disc corresponds to $b=0^\circ$). 
We calculate the minor axis of central galaxies using
Eq. \eqref{eq:tensor}, where the sum is over all the stellar
particles within a radius of 10\kpc{} from the centre of a
galaxy. The minor axis can be robustly determined since the majority
(99\%) of galaxies have minor to major axes ratio, $c/a$, less than
0.9. We show results for our full sample of MW-mass galaxies, which
consists of both disc and spheroid morphologies.

We find that, on average,
$\uvec{n}_{\rm orbit}$ is preferentially aligned with the minor axis
of the stellar distribution, $\uvec{e}_{\rm 3; \ cen}$, with a median
angle of 41.3$^\circ$. The alignment is even
stronger for the \MWorbit{} subsample; however, the presence of an LMC
reduces this alignment, as shown by the blue dotted line. Due to the
small number of \MWorbit{} systems with an LMC-mass satellite (there
are 30 such objects), we cannot exclude that the differences between
the red solid and blue dotted lines in \reffig{fig:Theta_a8_cen} are
due to stochastic effects; a KS-test finds that
the two curves are consistent at the $1.8\sigma$ level.

\reffig{fig:Theta_a8_cen} illustrates that the satellite orbits are
preferentially in the plane of the central galaxy disc
\citep[e.g. see][]{Lovell2011,Cautun2015b} and that this alignment is
even stronger for \MWorbit{} systems, in which the majority of
satellites have co-planar orbits. As we have seen from
\reffig{fig:Sat_halo}, the \MWorbit{} systems are also the ones most
strongly aligned with the halo minor axis. When taken together, it
suggests that satellites with co-planar orbits are preferentially
found in systems in which the minor axes of the stellar disc and the
DM halo are well aligned. Such configurations correspond to systems in
which the directions of anisotropic infall have been roughly constant
over time, since, on average, the orientation of the stellar component
is determined by the early filaments along which gas was accreted
while the orientation of the DM halo is determined by late time
filaments \citep[e.g. see][]{Vera-Ciro2011,Wang2011a}.

The same argument also explains why we would expect systems with a
massive satellite to have a higher degree of misalignment between
their satellite distribution and their central galaxies, as seen when
comparing the \MWorbit{} and \MWorbit{} with LMC samples in
\reffig{fig:Theta_a8_cen}. A more massive satellite indicates a later
assembly of the host halo \citep{Amorisco2017} and thus a larger time
span between when most stars were formed and when the satellites were
accreted. This increases the chance that the early filaments along
which gas was accreted are misaligned with the late time filaments
along which satellites fall into the system.

The MW, with a satellite system which is perpendicular to the stellar
disc, is an outlier when compared to the typical \eagle{} system (see
\reffig{fig:Theta_a8_cen}). Nonetheless, we do find \eagle{} examples that
have the same satellites--stellar disc geometry as the MW. To assess
 how atypical the MW satellite system is, we define perpendicular
 configurations as the ones for which $\cos\theta\leq0.2$. There are
 5 out of 139 (${\sim}4\%$) such perpendicular configurations in the
 \MWorbit{} sample and 3 out of 30 (${\sim}10\%$) in the \MWorbit{}
 with LMC-mass satellite sample. Thus, the MW satellites--stellar
 disc configuration is rather unusual, but less so when
 accounting for the fact that the MW has a very bright satellite. In
 \refsec{sect:MW-like_examples} we study in more detail the 5
 \MWorbit{} systems that most closely resemble our Galactic satellite
 distribution and investigate their formation history in detail.

\subsection{The alignment of satellite systems with the surrounding large-scale structure}
\label{subsec:results:sat-LSS_alignment}

As discussed previously, anisotropic infall is one of the driving
factors behind the formation of flattened and rotating satellite
distributions
\citep[e.g.][]{Libeskind2005,Libeskind2011,Libeskind2014,Deason2011,Lovell2011,Shao2018a}. The
same process, anisotropic accretion of DM and gas, is responsible, at
least partially, for the alignments between the DM, gas and satellite
distribution studied in the previous subsections, and it further
implies that these components are preferentially aligned with the
large-scale structures (LSS) in which they are embedded
\citep[e.g.][]{Tempel2015,Velliscig2015b,Welker2015,Shao2016,GaneshaiahVeena2018,GaneshaiahVeena2019}. This
motivates us to study the alignment between satellite systems and the
surrounding LSS, and compare it with Galactic observations.

\begin{figure}
	\plotone{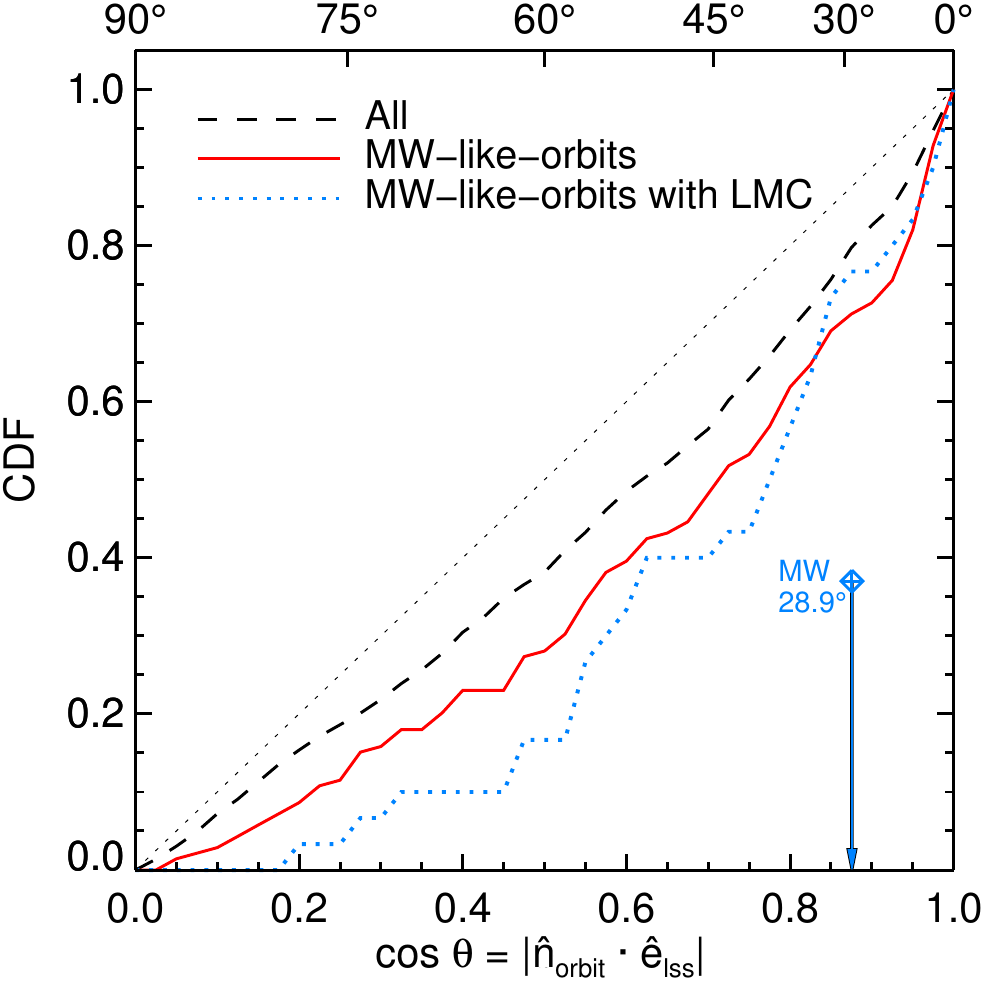}
	\caption{The CDF of the alignment angle, $\cos \theta$,=
          between the common orbital pole of satellites and, the
          normal, $\uvec{e}_{\rm lss}$, to the large-scale sheet in
          which the system is embedded. We show results for three
          samples: the full population of MW-mass systems (dashed black),
          the \MWorbit{} systems (solid red), and the \MWorbit{}
          systems that also have an LMC-mass satellite (dotted
          blue). The arrow indicates the measurement for our galaxy,
          which we obtained using the large-scale structure directions
          provided by \citet{Libeskind2015}. 
	}
	\label{fig:Theta_lss}
\end{figure}

We determine the orientation of the LSS using the NEXUS+ algorithm
\citep[][for a comparison with other cosmic web finders see
\citealt{Libeskind2018}]{Cautun2013,Cautun2014c}. This is a
multi-scale method that naturally determines the scale at which the
mass distribution is most anisotropic and that automatically
determines cosmic web environments, such as nodes, filaments and
walls. NEXUS+ takes as input the total matter density field smoothed
on a range of scales using a Gaussian filter. For each smoothing
scale, the method calculates the Hessian matrix of the smoothed
density field and, using its eigenvalues, determines the degree of
anisotropy of the mass distribution. At each location, NEXUS+ selects
the smoothing scale with the largest degree of anisotropy and the
eigenvectors of the Hessian matrix calculated for that smoothing scale
are then used to define the LSS directions. Here, we study the
alignment relative to the first direction of LSS collapse, which we
denote with $\uvec{e}_{\rm LSS}$. This direction is given by the
eigenvector corresponding to the largest eigenvalue and determines
the normal to the LSS sheet in which a system is embedded.

\reffig{fig:Theta_lss} shows the alignment between the satellite
distribution, characterised in terms of $\uvec{n}_{\rm orbit}$, and
the LSS direction, $\uvec{e}_{\rm LSS}$. We find a weak alignment
between the two orientations with a misalignment angle of $51.6^{\circ}$
(see \reftab{tab:median_alignment_angle}). It illustrates that the satellites
orbit preferentially within the plane defined by the LSS sheet
surrounding each system. The \MWorbit{} systems show an even better
alignment with the LSS than the full population. Furthermore, the
subsample with LMC-mass satellites shows a hint of an even stronger
alignment, but that sample is too small to arrive at statistically
robust conclusions. The weak present day alignment between the
satellite distribution and the LSS orientation is to be expected. This
alignment is largest when calculated at the time of infall of the
satellites \citep{Libeskind2014,Shao2018a}, and is weakened by the
subsequent evolution and re-arrangement of the cosmic web around each
halo \citep[e.g.][]{Vera-Ciro2011,Cautun2014c}.

To compare with the MW, we have calculated the angle between the MW
common orbital plane and the normal to the LSS as found by
\cite{Libeskind2015}. The latter was calculated using the
reconstructed velocity shear tensor in the Local Universe. We find
that the MW satellite distribution has a $29^\circ$ misalignment angle
with respect to the local LSS sheet, in qualitative
agreement with our theoretical predictions.
van Leeuwen (in prep.) has studied the cosmic web around
Local Group-like objects to find that $\uvec{e}_{\rm LSS}$ is
typically determined by the mass distribution on $2\Mpc{}$ scales
and that $\uvec{e}_{\rm LSS}$ shows a strong tendency to be
perpendicular on the direction connecting the two Local Group
members.

\section{The structure and formation history of the Galactic DM halo}
\label{sect:MW-like_examples}

As we discussed in the introduction, the MW classical satellites have
several properties that make them atypical of galactic satellite
systems in a $\Lambda$CDM universe. Previous studies have invoked such
features a potential tensions with the standard cosmological model
\citep[e.g.][]{Ibata2014b,Pawlowski2014c,Cautun2015}. However, while
the Galactic disc of satellites is rare, it is not rare enough to pose
a serious challenge to the $\Lambda$CDM model \citep[for details
see][and in particular their discussion of the look-elsewhere
effect]{Cautun2015}. Here, we take a different approach. We assume
that $\Lambda$CDM is the correct cosmological model and pose the
question: what do the atypical features of the MW satellite
distribution reveal about the structure and formation history of our
DM halo?

\begin{table*}
    \centering
    \caption{Selected properties of the 5 systems in the \eagle{}
      simulation that have similar satellite distributions to the MW's. The
      systems were chosen to have 8 of the 11 brightest satellites
      orbiting within a cone of opening angle, $\alpha_8<35^\circ$,
      and in a common orbital plane close to perpendicular ($\theta
      >78^\circ$) to the stellar disc of the central galaxy. The
      columns are as follows: (1)~system label, (2)~halo mass, (3)~halo
      radius, (4)~stellar mass, (5)~the angle, $\theta_{\rm
        orbit-halo}$, between the common satellite orbital plane and
      the halo minor axis, (6)~the angle, $\theta_{\rm orbit-cen}$,
      between the common satellite orbital plane and the central
      galaxy minor axis, (7)~the angle, $\theta_{\rm halo-cen}$,
      between the minor axes of the DM halo and central galaxy, and
      (8)~the stellar mass of the LMC-analogue if the system has one.
    }
   
    \renewcommand{\arraystretch}{1.1} 
    \begin{tabular}{ l l l l l l l l  } 
        \hline\hline
        Label & $M_{200}$ & $R_{200}$ & $M_{\star}$ & $\theta_{\rm orbit-halo}$ & $\theta_{\rm orbit-cen}$ & $\theta_{\rm halo-cen}$ & $M_{\rm\star \ LMC}$ \\
         & $[10^{10}\Msun]$ & $[\kpc]$ & $[10^{10}\Msun]$ & [deg] & [deg] & [deg] & $[10^{9}\Msun]$ \\
        \hline
        MW1 & 125.2 & 227.2 & 2.57 & $2.0$  & $72.9$ & 70.9 & 1.8 \\
        MW2 & 97.8  & 209.2 & 3.38 & $5.3$  & $80.4$ & 78.9 & -- \\
        MW3 & 87.3  & 201.4 & 1.82 & $27.9$ & $88.9$ & 89.0 & 4.1 \\
        MW4 & 76.0  & 192.4 & 0.98 & $9.8$  & $88.3$ & 82.5 & 2.7 \\
        MW5 & 37.4  & 151.9 & 0.64 & $31.3$ & $88.5$ & 67.0 & -- \\
        \hline\hline
    \end{tabular}
    
    \renewcommand{\arraystretch}{1.0}
    \label{tab:MW_analogues_properties}
\end{table*}

We identify MW analogues in the \eagle{} simulation that share the two
characteristics of the satellite distribution that stand out the most:
i) that 8 of the 11 classical satellites have nearly co-planar
orbits, and ii) that the common orbital plane of those
satellites is perpendicular to the stellar disc. These criteria
correspond to selecting from the \MWorbit{} subset the systems for
which the satellite distribution is nearly perpendicular to the
central disc (see Sec.~\ref{fig:Theta_a8_cen}), which we define as
having a misalignment angle, $\theta>78^\circ$ (i.e.
$\cos\theta< 0.2$). We find five such systems, which we label MW1 to
MW5, and whose properties are summarised in
\reftab{tab:MW_analogues_properties}.

\subsection{The structure of the DM halo}
\begin{figure}
    \centering
	\plotone{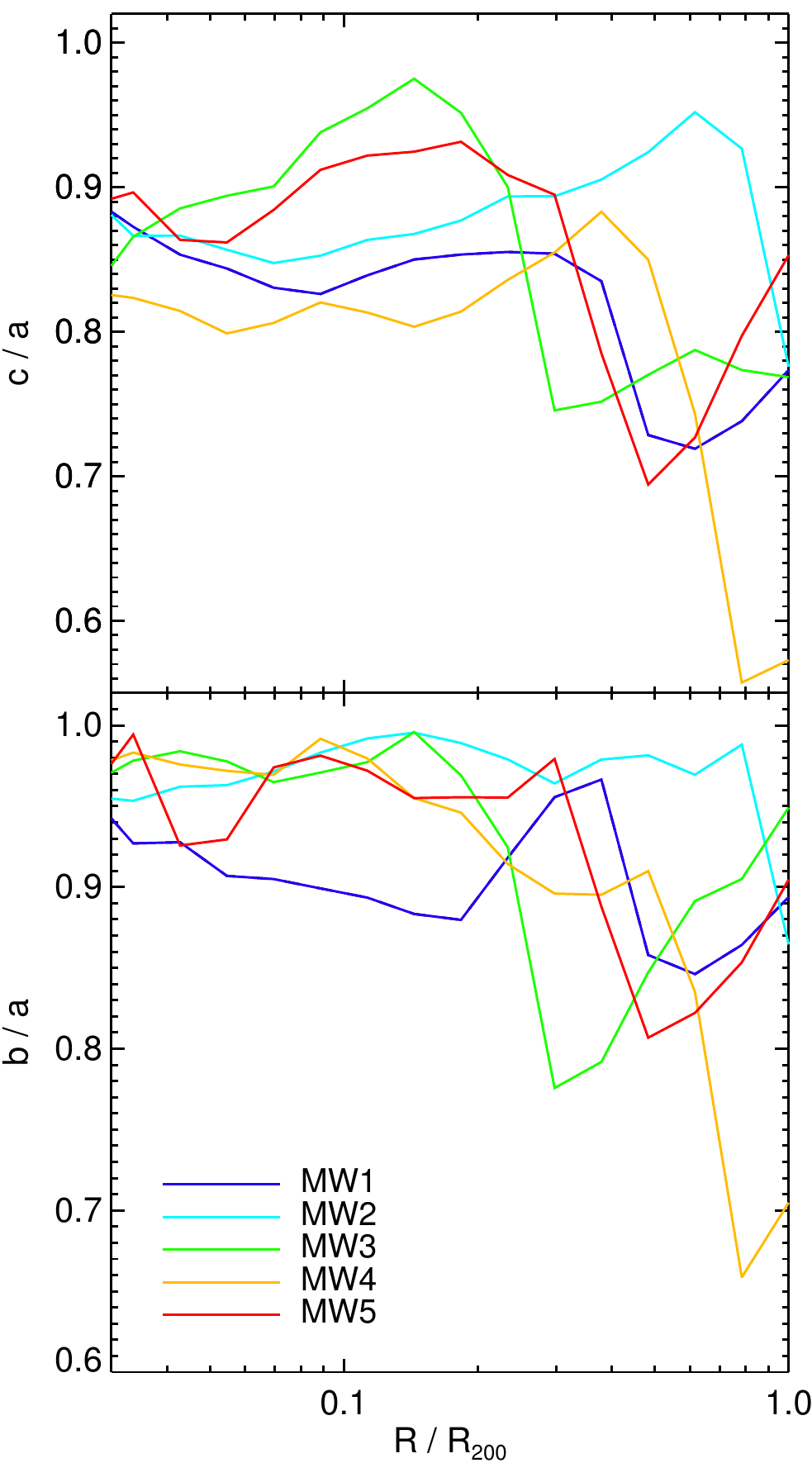}
	\caption{The $z=0$ axes ratios, $c/a$ (top panel) and $b/a$
          (bottom panel), for the five MW-analogues that have
          satellite distributions similar to the MW system. The axes
          ratios are shown as a function of the radial distance from
          the halo centre, normalised by the halo radius,
          $R_{200}$. Each point corresponds to the shape of the DM
          particle distribution enclosed in a sphere of the given
          radius.}
	\label{fig:MW_halo_ca}
\end{figure}

We start by studying the shape of the DM halo of our MW analogues as a
function of the distance from the halo centre, as illustrated in
\reffig{fig:MW_halo_ca}. For each radial bin, we calculate the shape
of the mass distribution within that radius. The inner regions of the
halo are only slightly flattened, with $b/a{\sim}0.95$ and
$c/a{\sim}0.85$, and the axes ratios show very little variation with
radius; we can therefore make robust predictions for the shape of the
inner DM halo. We note that the inner haloes in simulations that
include baryons are typically rounder than in DM-only
simulations \citep{Bailin_2005,Velliscig2015a,Chua2019,Prada2019},
with the dominant effect being the potential of the baryons, which is
very important for $r/R_{200}\leq0.2$. At larger distances, the halos
become systematically more flattened and, at the same time, show
greater halo-to-halo variation.

\begin{figure}
    \centering
	\plotone{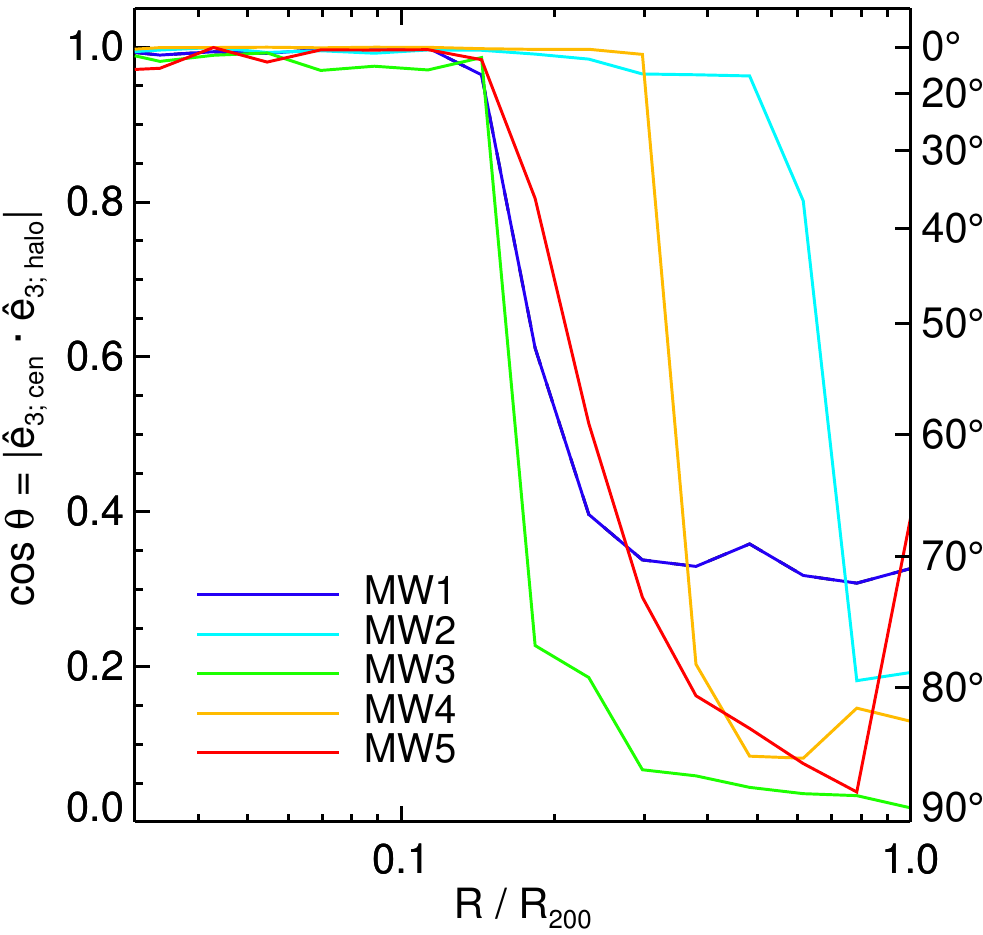}
	\caption{The alignment angle, $\cos \theta$, between the
          minor axes of the central stellar disc and of the DM
          halo. The halo shape is calculated as a function of radial
          distance. Each curve shows one of our five MW-analogues. The
          twist in the halo orientation, which is visible as a rapid
          change in the alignment angle, reflects the fact that that
          the outer halo is aligned with the satellite distribution,
          which is perpendicular on the central disc.
	}
	\label{fig:MW_ch}
\end{figure}

\begin{figure*}
    \centering
	    \includegraphics[width=.6\linewidth]{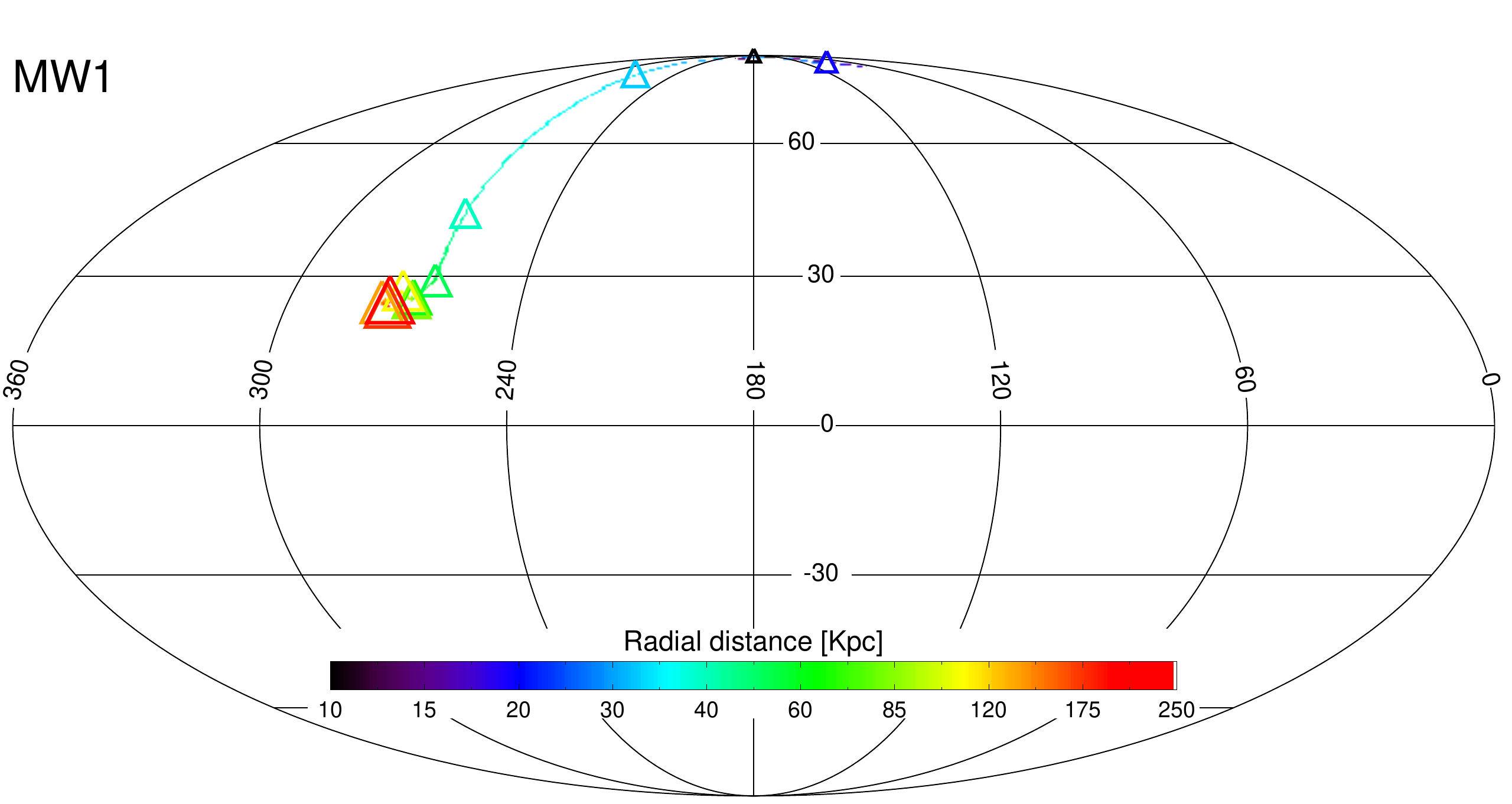}
	    \includegraphics[width=.3\linewidth]{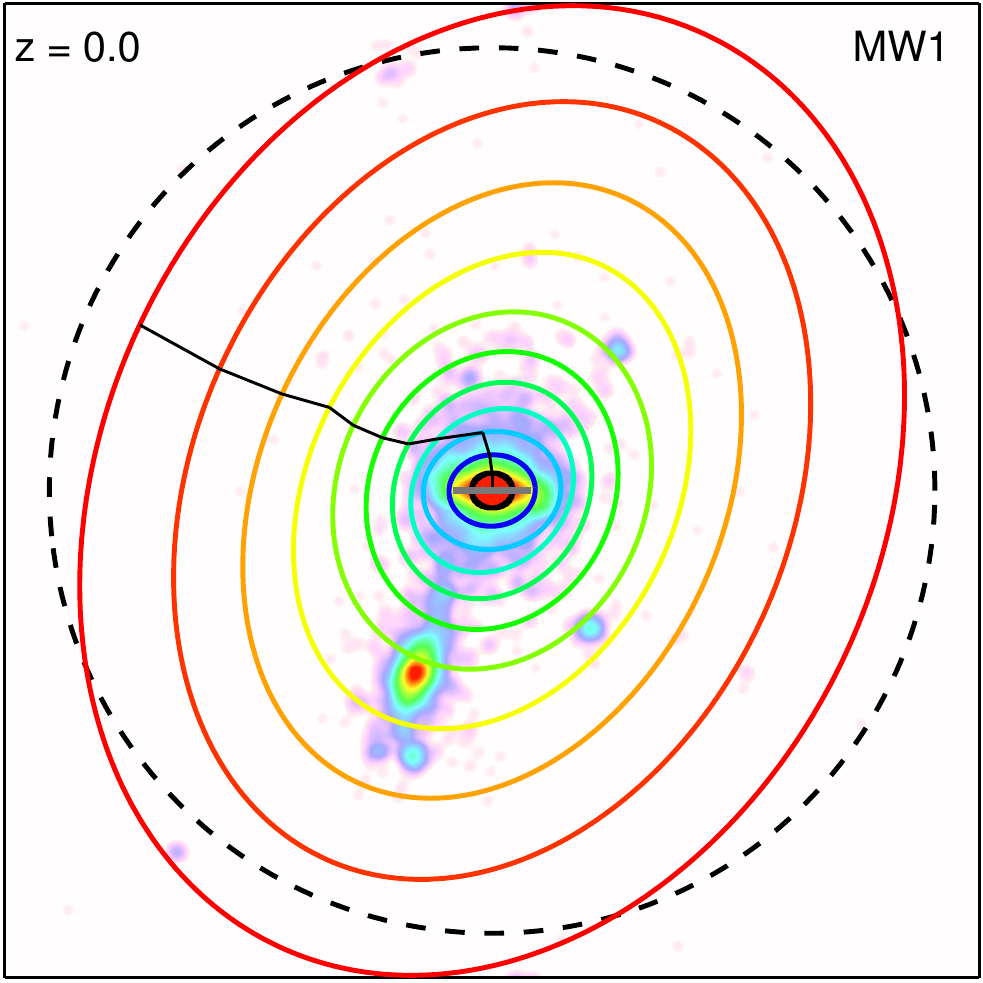}\\
		\includegraphics[width=.6\linewidth]{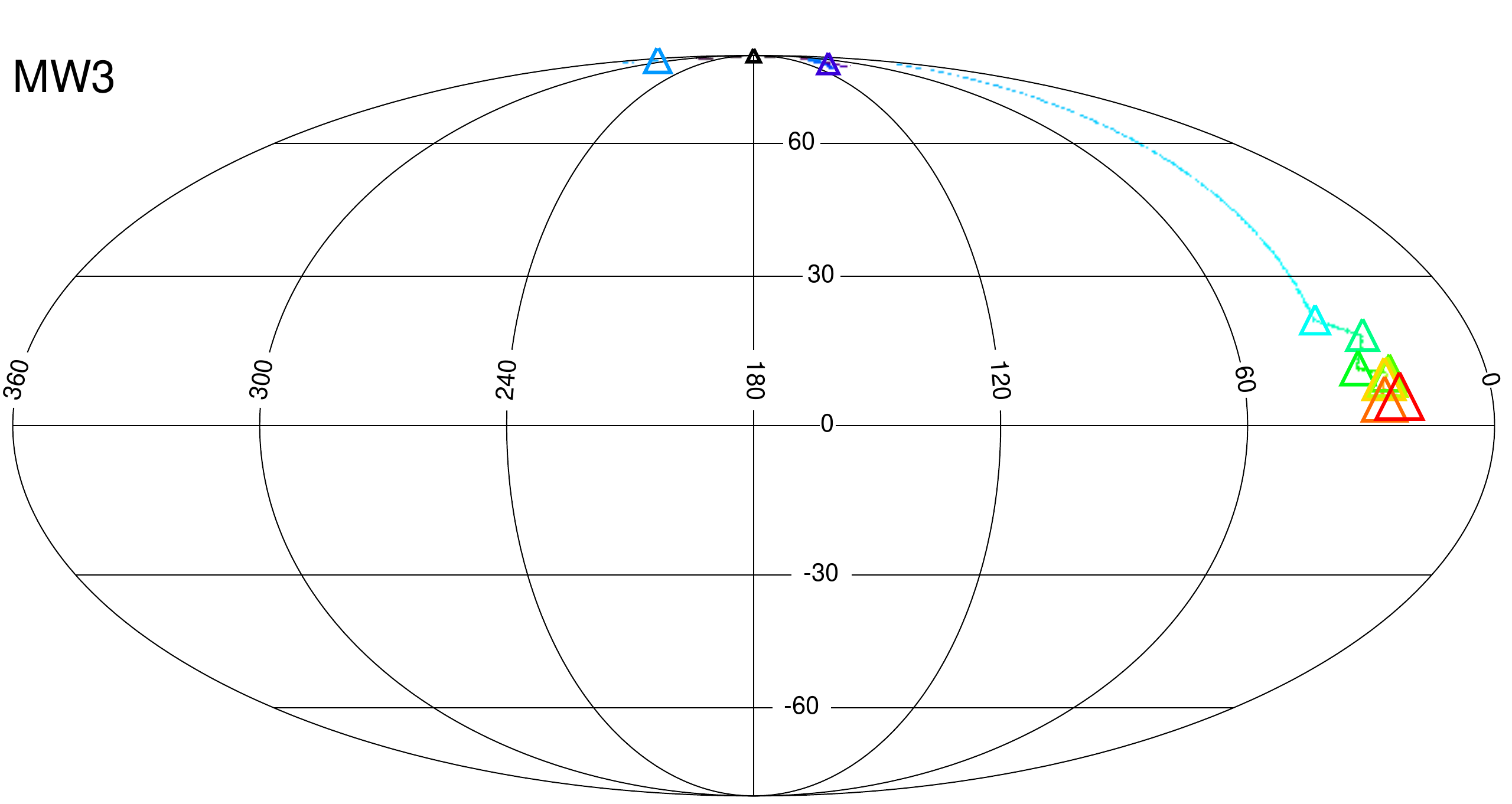}
	    \includegraphics[width=.3\linewidth]{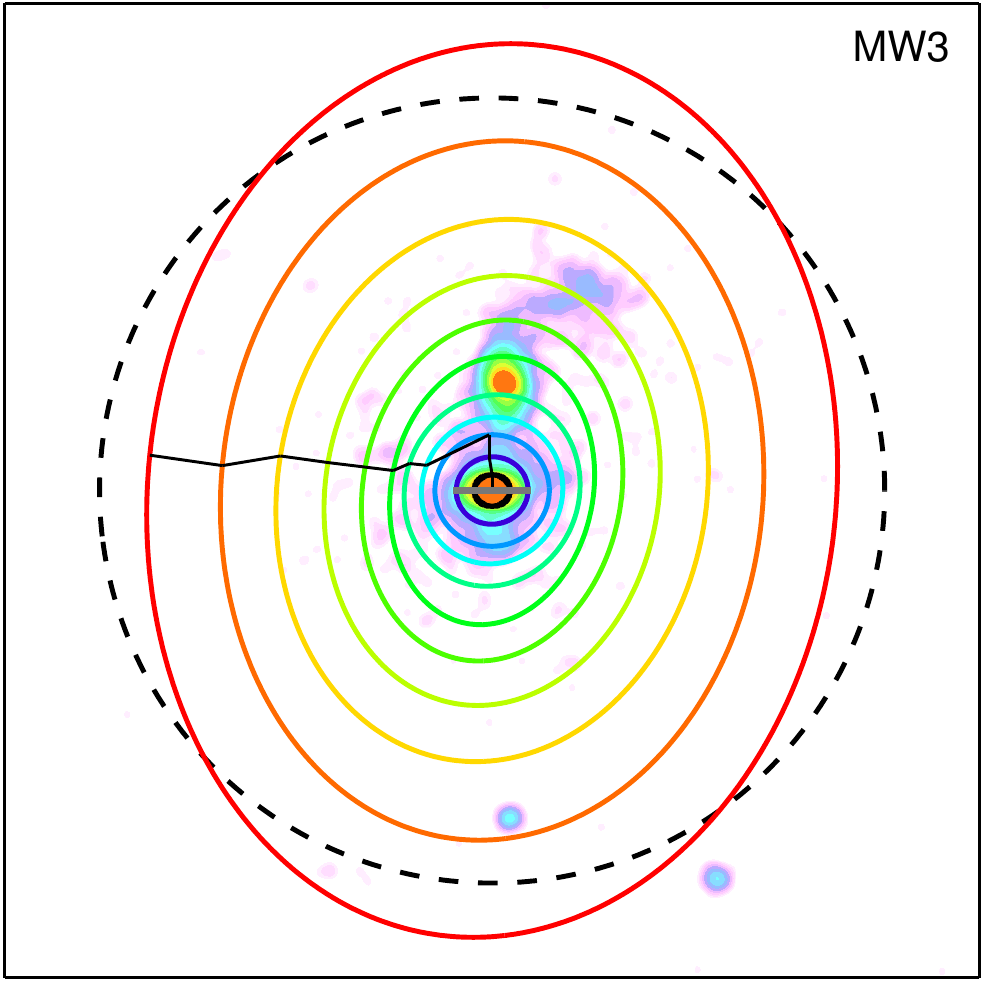}\\
	   	\includegraphics[width=.6\linewidth]{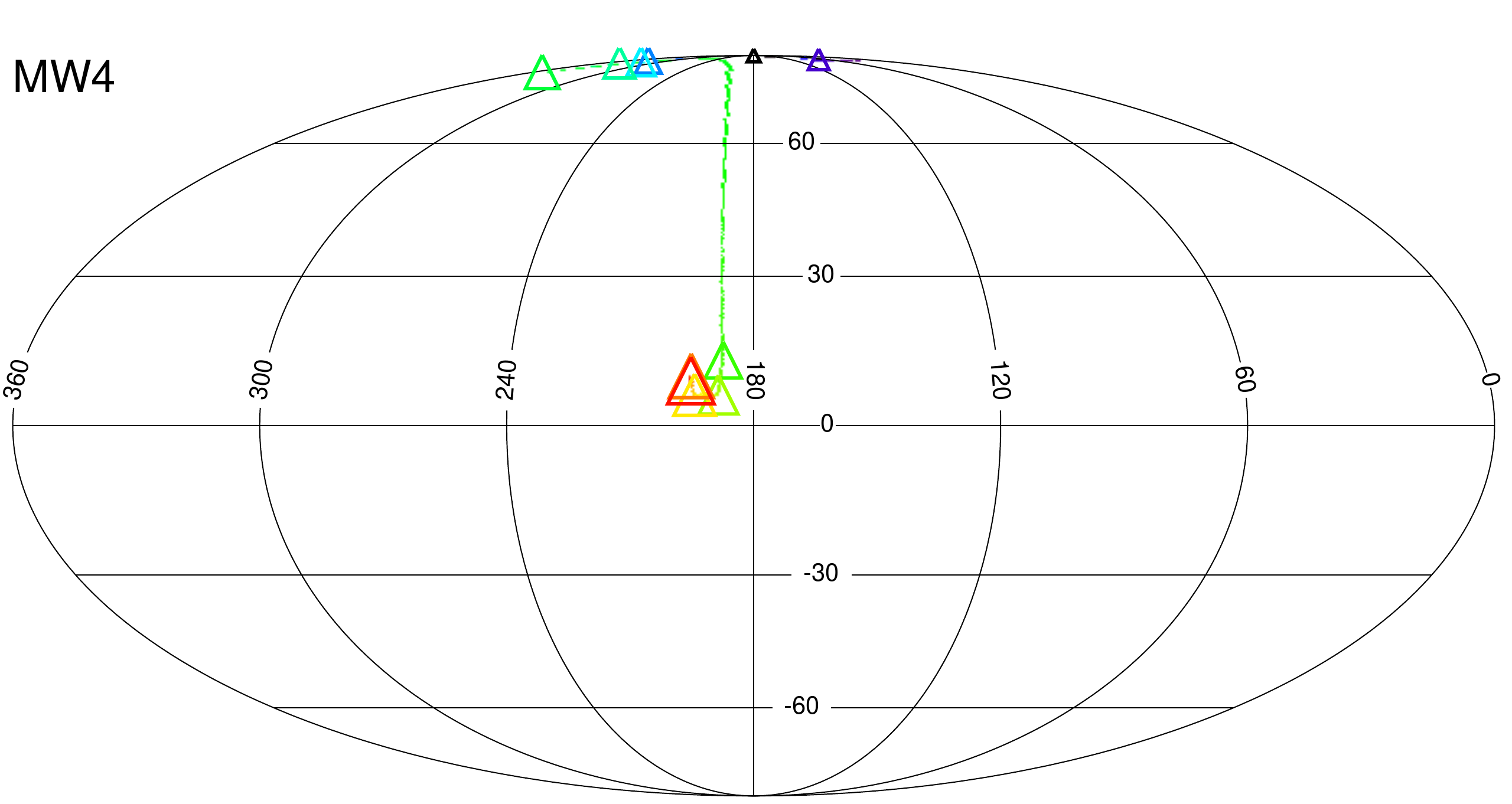}
    	\includegraphics[width=.3\linewidth]{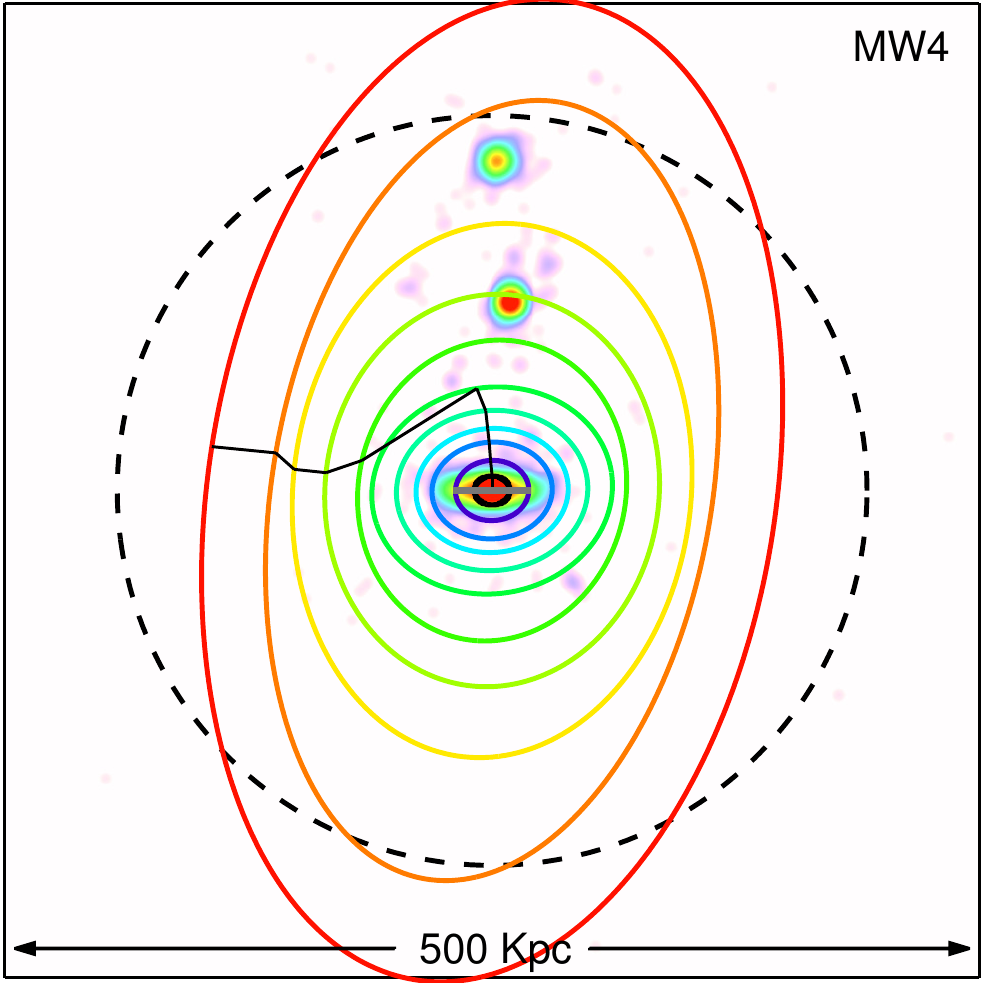}\\
	\caption{ \textit{Left panel:} Aitoff projection showing
	the orientation of the minor axis (triangles) of the host DM halo of the
	three MW analogues that have an LMC-mass satellite (MW1, MW3, and MW4).
	We measure the halo shape within spherical regions with radii between
	$10\kpc$ and $R_{200}$ (this is different for each host, see
	\reftab{tab:MW_analogues_properties}); the colours indicate the radii
	as shown in the legend. The coordinate system is given by the central
	galaxy stellar disc, with the disc being located in the	$b=0^\circ$ plane.
	\textit{Right panel:} the shape and orientation of the DM haloes at
	different radii. As for the left panel, the colours	indicate the radius.
	The background image shows the distribution of stars with the central
	galaxy seen edge-on along the x-axis and the rotating satellite distribution
	seen edge-on along the y-axis. The main axes of each ellipse are given
	by the major and the minor axis of the DM distribution within each 3D radius.
	Each ellipse is oriented such that it makes the same angle with the x-axis
	(i.e. the disc of the central galaxy) as the 3D angle between the minor
	axis of the halo and the stellar disc. For clarity, the position of the
	minor axis is highlighted by the solid black line that connects the various
	ellipses. The black dashed line shows the halo radius, $R_{200}$. 
	}
	\label{fig:MW_shape}
\end{figure*}

We next examine how the orientation of the DM halo changes as a
function of radial distance. This is illustrated in
\reffig{fig:MW_ch}, which shows the alignment between the minor axes
of the central galaxy and the DM halo. The inner halo is very well
aligned with the stellar distribution, as seen in other galaxy
formation simulations \citep[see
also][]{Bailin_2005,Velliscig2015a,Tenneti2014,Shao2016}. But, at
farther distances, we see a very rapid shift in the DM halo
orientation, which changes by more than $70^\circ$ over a very narrow
radial range. We refer to this feature as the ``twist" of the DM halo.
The exact radius where the twist takes place varies from system to
system, but the existence of such a twist is a robust feature across
all our MW analogues. At even larger distances, the halo orientation
remains fairly stable and nearly perpendicular to DM distribution in
the inner region.

We have checked that the halo ``twist" seen in
\reffig{fig:MW_ch} does not depend on the definition used to
calculate the halo shape. A similar twist is seen if instead we were
to use the reduced mass tensor \citep{Frenk1988,Dubinski_1991}, with
the only difference being that the twist is not as sharp. This
difference is due to the reduced tensor giving equal weight to all
DM particles, so we need to include more particles (i.e. go to
larger radii) outside the twist radius to see it. Similarly, in the
three cases that have an LMC-mass satellite, the twist is not
determined by the LMC's DM particles. We checked this by removing
all particles associated with the LMC-mass satellite before infall
into their MW-mass host. The orientation of the halo and the twist
radius hardly change when removing the DM particles associated to
the LMC-mass satellite.

In \reffig{fig:MW_shape} we present a more intuitive way of
visualising the orientation of the DM halo for our three MW analogues
(MW1, MW3 and MW4) that have an LMC-mass satellite. The left-hand
panels show the minor axis of the halo calculated at various radial
distances. The sky coordinates are fixed according to the stellar
distribution of each central galaxy, with the plane of the disc
corresponding to $b=0^\circ$. The sky projection clearly shows the
twist of the DM halo: the minor axis of the halo, which is found at
$b{\sim}90^\circ$ in the inner regions, undergoes a rapid change to
$b{\sim}0^\circ$ at large radial distances. The panels also illustrate
that the minor axis orientation generally varies by ${\sim}10^\circ$
or less between neighbouring bins, a signature of a smooth change in
the different directions along which the halo assembled
\citep{Vera-Ciro2011}. However, the halo ``twist" represents a
dramatic change in orientation, with the minor axis varying by
${\sim}70^\circ$ from one radial bin to the next. This suggests that,
to a good approximation, these DM haloes can be modelled as an inner
component with minor axis at $b=90^\circ$ and an outer component with
minor axis at $b{\sim}0^\circ$. 

The right-hand panels of \reffig{fig:MW_shape} illustrate both the
shape and orientation in projection of the DM halo for the MW1, MW3
and MW4 systems. To highlight its relation to the stellar distribution,
we select a Cartesian coordinate system in which the central galaxy
is seen edge-on along the x-axis and the rotating plane of satellites
is also found roughly edge-on but along the y-axis. Such geometries
are possible for our sample of MW-analogues since the rotating plane
of satellites is perpendicular to the central disc. In each panel,
the stellar distribution is shown by the background colours, with the
LMC-mass analogue being clearly visible as a massive blob. The halo
shape and orientation are represented by ellipses, with each ellipse
corresponding to the DM distribution in a sphere centred on the central
galaxy. The axes of each ellipse are given by the major and minor axes
of the 3D DM distribution, and the ellipse is orientated to make the
same angle with the x-axis (i.e. the central stellar disc) as the 3D
angle between the halo minor axis and the stellar disc. The figure
provides a compelling illustration of the complex geometry that
characterises our MW analogues.

The ubiquity of a ``twist" in all our MW analogues suggests that such
a feature ought to be present in our Galactic DM halo
too. Unfortunately, the satellite distribution cannot constrain the
exact radius where the twist would happen, which for our sample varies
from $30\kpc$ for MW3 to $150\kpc$ for MW2. Evidence for a twist in
the Galactic halo has been claimed before when modelling the orbit of
the Sagittarius stream (see the interpretation of the
\citealt{Law2010} MW model by \citealt{Vera-Ciro2013}), but the
validity of this claim has been hotly debated, especially because the
massive DM halo in which the LMC resides could introduce systematic
effects \citep[e.g. see][]{Vera-Ciro2013,Gomez2015}. If we take the
results of the \citeauthor{Law2010} Galactic model at face value, then
the Galactic halo twist must be inside the orbit traced by
Sagittarius, potentially as close as a few tens of kiloparsecs from
the Galactic Centre.

\begin{figure}
    \vspace{0.5cm}
    \centering
	\plotone{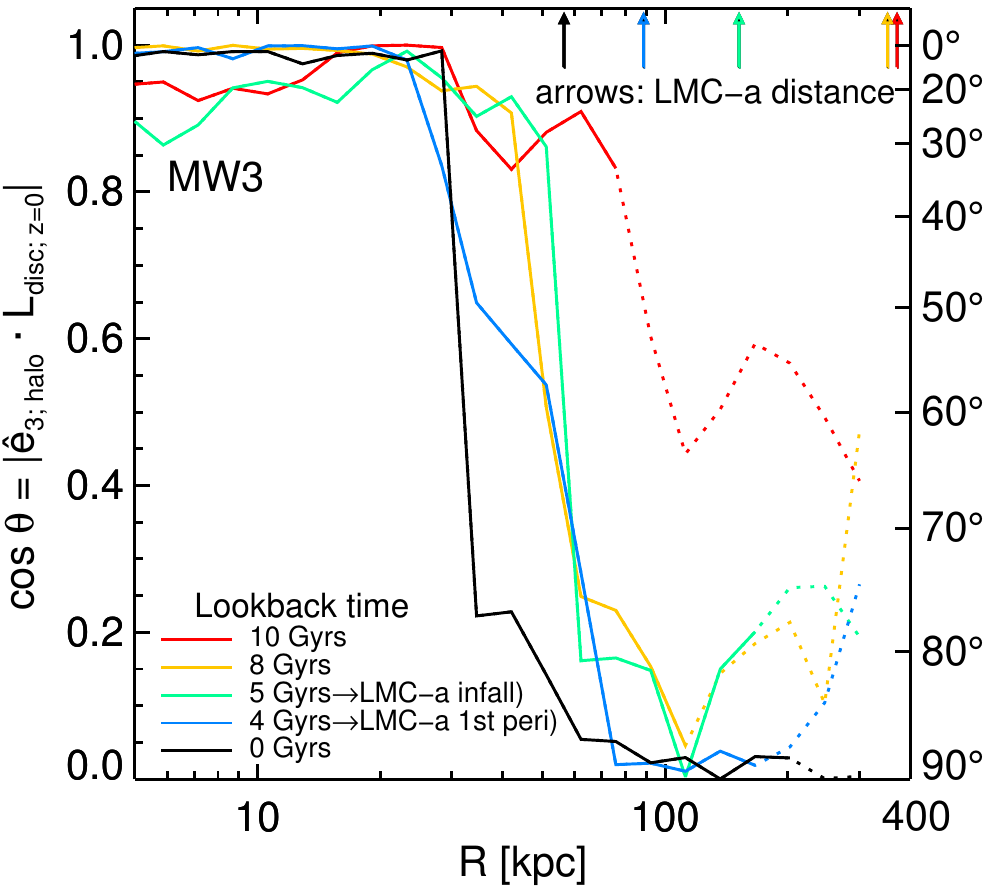}
	\caption{Time evolution of the alignment angle
            between the central galaxy's angular momentum and the
            minor axis of the DM halo for the MW-analogue
            MW3. As in \reffig{fig:MW_ch}, the halo shape is
            calculated as a function of distance from the centre,
            which is shown on the x-axis. The various lines show the
            system at lookback times of 10, 8, 5, 4, and 0~Gyrs. The
            shape of the matter distribution for the large R bins was
            calculated using also DM particles outside the halo radius
            and, to distinguish them, those results are shown by a
            dotted line. The vertical arrows indicate the radial
            distance of the LMC-analogue satellite.
	}
	\label{fig:MW3_minor_time}
\end{figure}

\subsection{The formation of a twisted halo}
\begin{figure}
    \centering
\plotone{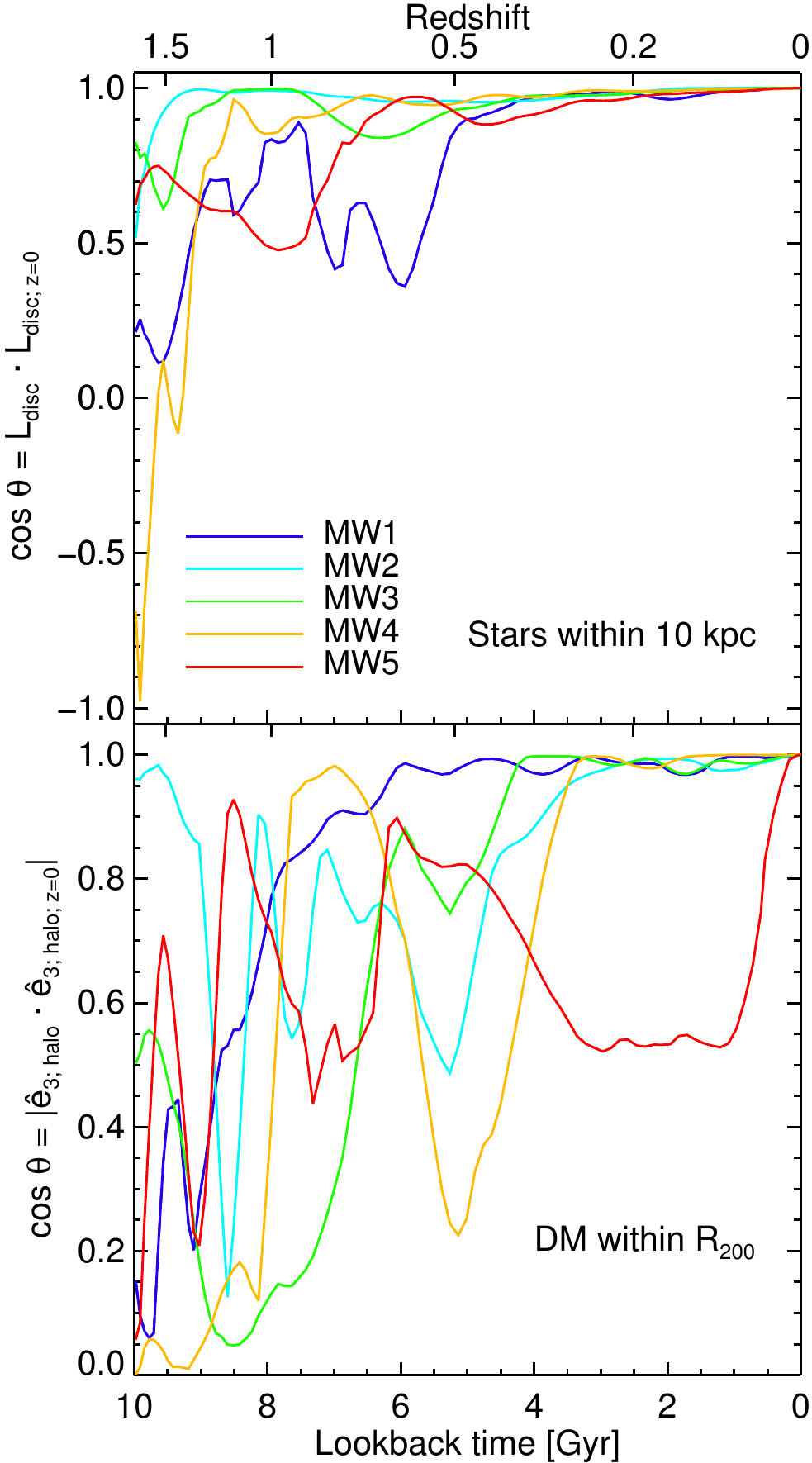}
\caption{\textit{Top panel:} the orientation of the central galaxy's
  angular momentum at different stages of the formation history. The
  orientation is relative to the direction of angular momentum at
  $z=0$. \textit{Bottom panel:} the same but for the minor axis of the
  whole DM halo. }
	\label{fig:MW_time}
\end{figure}

\begin{figure*}
    \centering
    \resizebox{0.61\hsize}{!}{\includegraphics{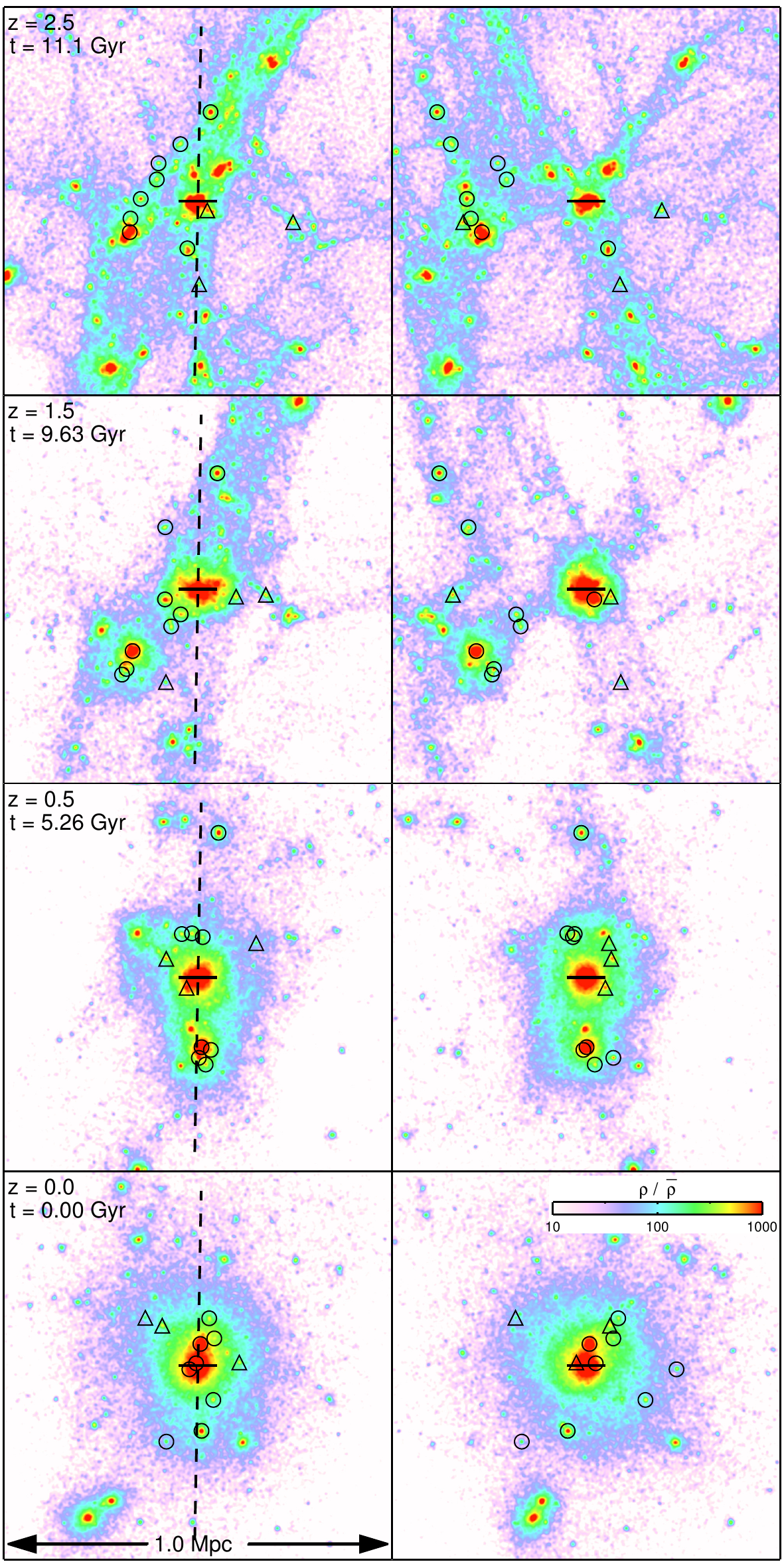}}
	\caption{ The evolution of the DM distribution within 0.5 physical
	Mpc around one of our MW analogues, MW3. The colours show the
	projected DM density, with red colours corresponding to high density
	regions and white corresponding to low density regions (see colour
	bar in the bottom-right panel). The rows show the system at redshifts:
	2.5, 1.5, 0.5, and 0.0, respectively. The left-hand column corresponds
	to the coordinate system in which the disc of the central galaxy at
	$z=0$ is seen edge-on along the x-axis (see horizontal black solid line)
	and the rotating plane of satellites at $z=0$ is seen also edge-on but
	along the y-axis (vertical black dashed line). The right-hand column
	shows the system after a $90^\circ$ rotation, with the central disc
	still seen edge-on along the x-axis but with the rotating plane of
	satellites seen face-on. The satellites and their progenitors are
	shown as open circles (for the 8 out of the 11 brightest satellites
	with co-planar orbits at $z=0$) and open triangles (for the remaining
	3 satellites).
	}
	\label{fig:DM_distribution}
\end{figure*}

We start by studying if the halo twist is a long-lived
feature. We would expect this to be the case since the presence of
the twist was inferred from the orbits of the classical satellites,
which have an orbital period of a few Gyrs. Indeed, this is
the case for all our five MW-analogues: the halo twist formed
${\sim}10$ Gyrs ago and, since then, has shown little change. We
highlight this in \reffig{fig:MW3_minor_time}, where, for the MW3
analogue, we plot halo orientation as a function of distance
from the centre at various lookback times. We find that the twist
started forming 10 Gyrs ago in the matter distribution just outside
the halo radius (see dotted red line in the figure). As the MW3 halo
accreted that material, the twist radius shifted inwards to
${\sim}50\kpc$ and remained at that position until 4 Gyrs ago. That
is when the MW3 halo experiences the first pericentre passage of an LMC-mass
satellite; this shifted the twist inwards to a distance of $30\kpc$
from the centre, where it has remained until the present day. Three of
our five MW-analogues experience an inward shift of the halo twist
radius: two of the three systems with an LMC-mass satellite and one
of the two systems without an LMC-analogue.

\reffig{fig:MW3_minor_time} also highlights that after
formation the twist radius is rather sharp, with the exception of
the transient stage when the twist moves inwards. Moreover, the
twist formed prior to the infall of massive satellites that survive
to present day. For example, MW3 has a massive LMC-analogue
satellite that brought in $20\%$ of the $z=0$ total mass of the
system. In this case, the twist was fully formed already 8 Gyrs ago,
when the LMC-analogue was over 300\kpc{} from the host halo
centre. As we shall discuss next, the twist is associated with a
change in the orientation of the dominant cosmic web filament
feeding the system, which brings in a large amount of mass into the
growing halo.

We now investigate the effects that produce a twist in the DM haloes
in all our MW analogues. We focus on answering two questions:
i) is the twist produced because the spin of the stellar disc
flipped at some point as a result of either a merger with or a flyby
by other galaxies
\citep[e.g.][]{Bett2012b,Bett2016,Dubois2014,Earp2017}? or ii)
is the twist due to a variation in the direction of the (anisotropic)
infall of satellites?

To answer these questions, we follow the variation in time of the
orientation of the central galaxies and their DM halos in each of our
MW analogues. This is shown in \reffig{fig:MW_time} where we plot the
orientation relative to that at the present day. We find that the
orientation of the central discs has been relatively stable in the
past 5 Gyrs and potentially even longer for some systems, such as MW2
and MW4. During the last several gigayears, the discs experience only
minor changes in orientation \citep[see also][]{Earp2019}, typically
$\lesssim20^\circ$; these cannot explain the ${\sim}90^\circ$
misalignment between the stellar and DM components in our MW
analogues. At early times, the orientation of the stellar disc can
vary more rapidly (e.g. see MW1 in the top panel of
\reffig{fig:MW_time}) but this is typically the period when
the stellar mass was only a small fraction of today's value (see
\reffig{fig:MW_mass_assembly} in the Appendix).

In the bottom panel of \reffig{fig:MW_time} we study the changes in
the orientation of the DM halo. We see rather large variations even at
late times (e.g. MW5): the orientation of the halo is much less stable
in time than that of the stellar disc. The orientation of the halo is
affected by recently accreted material, which, being at large
distances from the halo centre, makes a large contribution to the
mass tensor. The distribution of recently accreted material is
determined by the geometry of the cosmic web surrounding the system
and, thus, the variation in halo orientation is a manifestation of
variations in the LSS within which it is embedded. We have confirmed
this point visually by viewing movies of the evolution of these
systems. They show that the filaments feeding the halo can vary in
time, especially between early ($z>1$) and late ($z<0.5$) epochs, and
that the variation is due to a mismatch between the small-scale cosmic
web which feeds the early growth of the halo and the slightly larger
scale web that is important for the late halo growth. The evolution
of one such system, MW3, is illustrated in \reffig{fig:DM_distribution},
which shows the DM distribution surrounding the halo from $z=2.5$ up
to present day. Note that, in general, the small- and large-scale webs
are well aligned \citep{Aragon-Calvo2007PhD,Rieder2013}. The small and
rather special sample of systems we are considering here are not
representative of the average $\Lambda$CDM halo.

We also checked that the sudden change in halo orientation is not due
to the accretion of the LMC-mass analogues found in the MW1, MW3 and
MW4 systems. In fact, the DM halo changes orientation before the LMC
analogue falls in (see \reffig{fig:LMC_orbits} in the Appendix
for the orbits of the LMC analogues). This is to be expected since
the most massive satellites are accreted along the most prominent
filament \citep{Shao2018a} and that is already in place before the infall
of the LMC analogue. It is this prominent filament,
along which the LMC analogue falls in, that determines the orientation
of the host halo.

\section{Conclusions}
\label{sect:conclusions}

We have analyzed analogues of the Milky Way in the \eagle{}
cosmological hydrodynamics simulation of galaxy formation in order to
learn about the likely structure, shape and orientation of the Milky
Way's dark matter halo. Our sample of MW analogues consists of
\eagle{} halos of mass $M_{200}{\sim}10^{12}\Msun$ whose brightest 11
satellites have similar spatial and kinematical properties to the
classical satellites of our Galaxy. In particular, we defined the
subset of ``\MWorbit{}'' systems as those in which the majority of the
satellites orbit in a single plane. This selection was motivated by
the observation of \citet[][see also
\citealt{Pawlowski2013}]{Shao2019} that 8 out of the 11 classical MW
satellites have orbital poles that lie within a very narrow,
$\alpha_8^{\rm MW}=22^\circ,$ opening angle. To obtain a reasonably
sized sample of counterparts in the relatively small volume of the
\eagle{} simulation [$(100 {\rm Mpc})^3$)] we relaxed the criterion
slightly, requiring that 8 of the brightest 11 satellites should have
orbital poles within an $\alpha_8=35^\circ$ opening angle.

From our subsample of \MWorbit{} DM haloes we conclude:
\\[-.55cm]
\begin{enumerate}
\item Halos that, like that of the MW, host co-planar satellite orbits
  tend to be more flattened (lower $c/a$ axis ratio) than the full
  population of haloes of similar mass. The \MWorbit{} systems also
  have $b/a$ ratios closer to unity than the full sample (see
  \reffig{fig:Pdf_ca_halo}).
  \\[-.3cm]
\item The normal to the common orbital plane of satellites is well aligned with the
  minor axis of the DM host halo. The alignment is even stronger for
  the \MWorbit{} subsample (see \reffig{fig:Sat_halo}).
  \\[-.3cm]
\item From these results, we predict that the minor axis of the actual
  MW DM halo should be pointing along the direction
  $(l,b)=(182^\circ,-2^\circ)$, with the 50, 75 and 90 percentile
  confidence intervals corresponding to angular uncertainties of
  17.3$^\circ$, 26.9$^\circ$, and 36.6$^\circ$ respectively
  (see \reffig{fig:MW_sky}).
  \\[-.3cm]
\item The common orbital plane of satellites in the simulations is
  preferentially aligned with the central stellar disc, but this
  alignment is not as strong as that with the DM halo (see
  \reffig{fig:Theta_a8_cen}).
  \\[-.3cm]
 \item The presence of an LMC-mass satellite does not affect
  the satellite orbital plane--DM halo alignment, but it weakens the
  satellite orbital plane--central disc alignment. 
 \\[-.3cm]
 \item The planes of satellites have only a weak
  alignment with the present day LSS environment in
  which they are embedded (see \reffig{fig:Theta_lss}).
 	 
\end{enumerate}

The MW satellite distribution has another unusual feature: the common
orbital plane (and the associated plane of satellites) is almost
perpendicular to the stellar disc. Such configurations are rare in the
\eagle{} simulation, where most satellites orbit in the plane of the
central galaxy. To understand the implications of this strange
perpendicular arrangement, we selected those \MWorbit{} systems in which the
majority of bright satellites orbit in the plane perpendicular to the
stellar disc. Only five such examples are to be found in \eagle{},
corresponding to ${\sim}4\%$ of the \MWorbit{} sample. Three out of
the five have an LMC-mass satellite indicating that the presence of a
massive satellite makes a perpendicular configuration between the
orbits of satellites and the central disc more likely.

From this subset of 5 MW-analogues, which represent the closest match
in \eagle{} to the spatial and kinematical distribution of classical
satellites in the MW, we find:
\\[-.55cm]
\begin{enumerate}
\item In the inner ${\sim}30\kpc$, the halos of the MW-analogues have
  axis ratios, $b/a=0.85$ and $c/a=0.95$, with little halo-to-halo
  variation. The outer parts of the halo are more flattened than the
  inner parts and show larger halo-to-halo variation (see
  \reffig{fig:MW_halo_ca}).
  \\[-.3cm]
\item The DM halo of each MW-analogue is ``twisted'' such that the
  orientation of the outer halo is perpendicular to that of the inner
  halo. Since the main plane of the inner halo is aligned with the
  central disc, the outer halo is nearly perpendicular to the stellar
  disc. The location of the twist varies amongst halos, but always
  occurs suddenly, in a very narrow radial range (see
  \reffig{fig:MW_ch}).
  \\[-.3cm]
\item In all our MW analogues, the twist is due to a shift in the
  direction of (anisotropic) accretion between early and late times,
  which is reflected in the different the orientations of the inner
  and outer DM halo. The central disc is quite stable once most of the
  stars have formed, at redshift, $z\lesssim0.5$ (or ${\sim}5$ Gyrs
  lookback time).
\end{enumerate}

Cosmological hydrodynamical simulations predict that the
central galaxy and the outer halo can be misaligned, with a median
angle of 33$^\circ$ \citep[e.g.][]{Bett2007,Tenneti2015,Velliscig2015a,
Shao2016}. However, only a small fraction of systems have a
${\sim}90^\circ$ misalignment, which is what we predict to be the
case for the MW. For example, only ${\sim}10\%$ of our MW-mass sample
consists of cases where the outer halo is close to perpendicular
(i.e. at an angle of 80$^\circ$ or higher) to the stellar disc.
The MW's misaligned DM halo is another feature, on top of the plane of
satellite galaxies, that makes our galaxy stand out.

The ``twisted" DM halo inferred for our Galaxy by our analysis is
consistent with the Galactic model proposed by \citet{Vera-Ciro2013}
in which the inner halo is aligned with the MW disc while the outer
halo is perpendicular to it. This model is based on the analysis of
the orbit of the Sagittarius stream by \citet{Law2010} who argued that
this requires the minor axis of the Galactic halo to be perpendicular
to the stellar disc. Furthermore, our prediction for the orientation
of the minor axis of the halo, $(l,b)=(182^\circ,-2^\circ)$, matches
very well the orientation inferred by \citeauthor{Law2010}\footnote{We
  applied a 180$^\circ$ shift in $l$ to the value reported by
  \citeauthor{Law2010} to account for the fact that we measure an
  orientation and not a vector (i.e. both vectors $\mathbf{x}$ and
  $-\mathbf{x}$ correspond to the same orientation).},
$(l,b)=(187^\circ,0^\circ)$. We note that the
halo orientation inferred by \citeauthor{Law2010}  is still a matter of
debate, largely because it ignores the gravitational influence of the
LMC \citep{Vera-Ciro2013,Gomez2015}, which is thought to be rather massive
\citep{Penarrubia2016,Laporte2018,Shao2018,Cautun2019} and this could
introduce systematic uncertainties. In a recent study, 
\citet{Erkal2019} have argued that even when including the LMC
potential, the orbit of the Orphan stream prefers an oblate Galactic
halo with minor axis pointing towards
$(l,b)=(176.2^\circ,-13.1^\circ)$; this agrees very well with our
own prediction for the orientation of the outer halo. Thus,
our study provides independent
and robust evidence that our Galactic DM halo is indeed ``twisted'', a
conclusion that could perhaps be tested further with GAIA data.

One of the limitations of our analysis is that, in order to obtain a
large sample of MW-like systems, we had to relax the criteria for
selecting satellite distributions with a majority of co-planar
satellite orbits. Our \MWorbit{} sample consists of systems where
eight satellites have orbital poles within opening angle,
$\alpha_8=35^\circ$, while for the MW the opening angle is
$\alpha_8^{\rm MW}=22^\circ$. We find that limiting our analysis to
systems with small $\alpha_8$ values leads to an even tighter
alignment between the normal to the common orbital plane of satellites and the halo
minor axis although with increased noise. Future simulations with much
larger volumes than \eagle{} will provide larger samples of systems
with small enough values pf $\alpha_8$, potentially enabling more
robust constrains on the orientation of the Galactic DM halo.

A larger sample of MW-analogues would be needed to investigate whether
the location of the twist can be inferred from the properties of the
satellite sample itself. For example, satellites accreted early have
fallen along different directions from satellites accreted later on, so
contrasting the orbits of early versus late accreted satellites could
constrain the lookback time at which the halo switched
orientation. The earlier the switch, the further in it happens.

All our \eagle{} MW-analogues exhibit a twisted DM halo and, on this
basis, we have argued, that this feature is a generic prediction of
$\Lambda$CDM. While twisted haloes have so far only been identified in
the \eagle{} simulation, we expect this feature to be independent of
the galaxy formation physics. The tight alignment between satellite
orbits and the outer DM halo is driven by gravitational collapse and
thus is largely insensitive to the details of baryonic
physics. Similarly, the tight alignment between the central galaxy and
the inner halo is a consequence of the DM in the inner regions
conforming to the gravitational potential which is dominated by the
baryonic distribution. Our simulations suggest that twisted DM haloes
should be commonplace in a $\Lambda$CDM universe.

\section*{Acknowledgements}
We thank the anonymous referee for detailed and thoughtful comments
that have helped us to improve the paper significantly.
We thank Jeremy Bailin, Vasily Belokurov, Denis Erkal,
and Amina Helmi for many useful comments and discussions.
SS, MC and CSF were supported by the European Research Council through
ERC Advanced Investigator grant, DMIDAS [GA 786910] to CSF. This work
was also supported by STFC Consolidated Grants for Astronomy at
Durham, ST/P000541/1 and ST/T000244/1. MC acknowledges support by the
EU Horizon 2020 research and innovation programme under a Marie
Sk{\l}odowska-Curie grant agreement 794474 (DancingGalaxies). AD is
supported by a Royal Society University Research Fellowship. This
work used the DiRAC Data Centric system at Durham University, operated
by the Institute for Computational Cosmology on behalf of the STFC
DiRAC HPC Facility (www.dirac.ac.uk). This equipment was funded by BIS
National E-infrastructure capital grants ST/P002293/1, ST/R002371/1
and ST/S002502/1, Durham University and STFC operations grant
ST/R000832/1. DiRAC is part of the National e-Infrastructure.

\section*{Data availability}
The \eagle{} data are publicly available at \url{http://icc.dur.ac.uk/Eagle/database.php}. The data produced in this paper are available upon reasonable request to the corresponding author.

\bibliographystyle{mnras}
\bibliography{bibliography}

\appendix
\section{Halo and galaxy mass accretion rates}
\label{sec:app_mass_growth}
\begin{figure}
    \centering
	\plotone{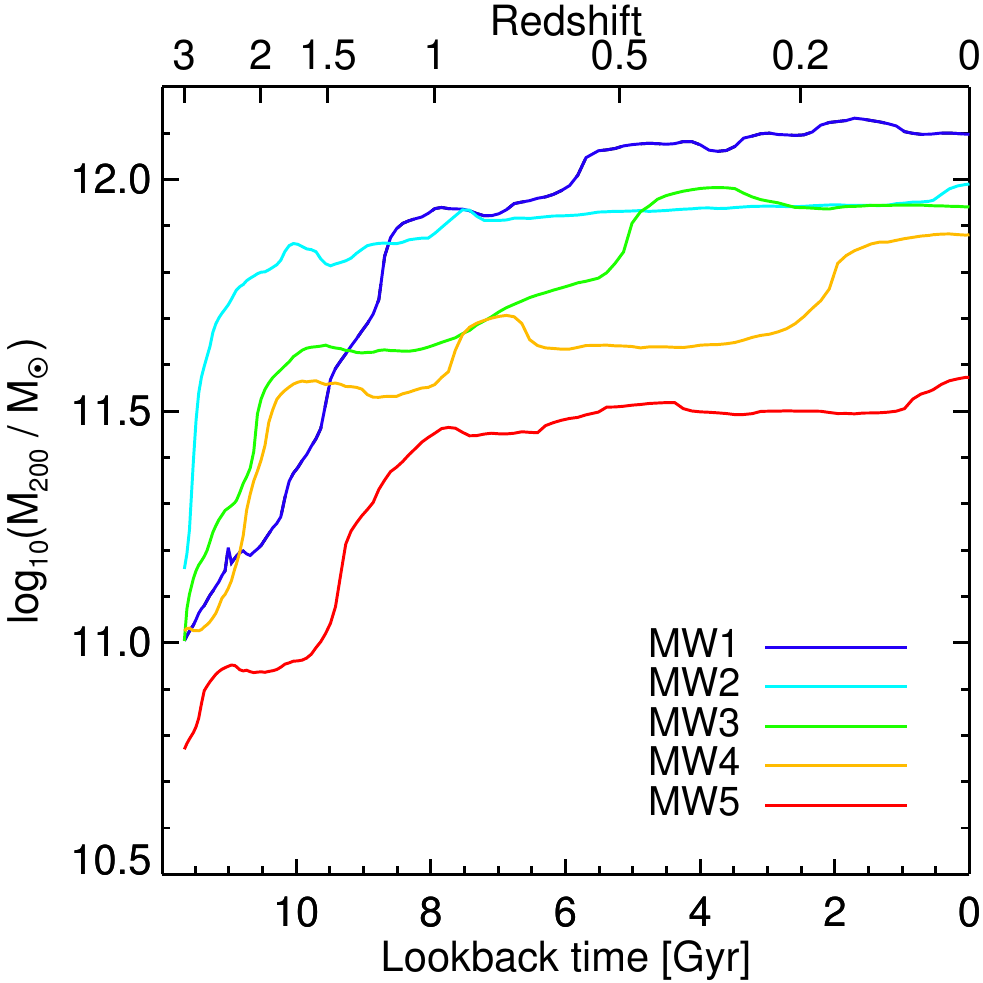}
	\plotone{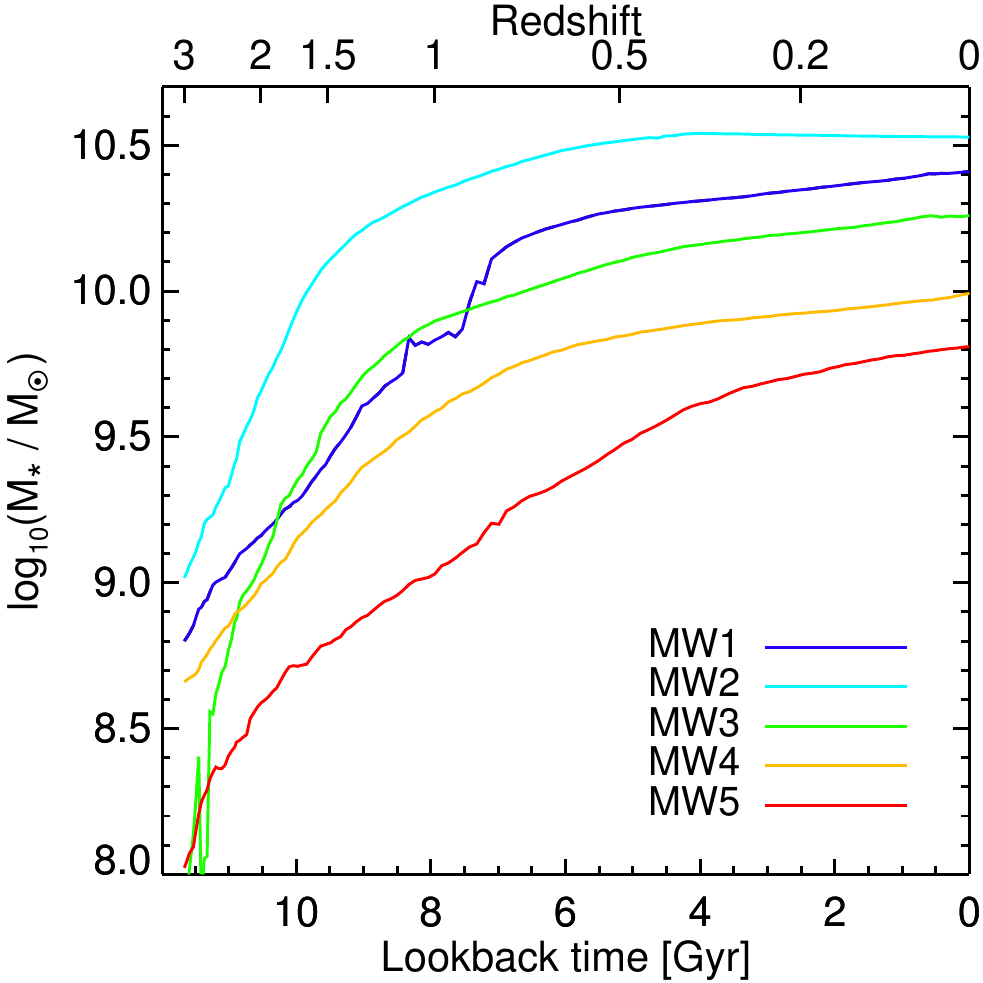}
	\caption{ The mass assembly history of the DM halo (top panel) and central galaxy (bottom
          panel) of the 5 MW analogues studied in detail in this paper. 
	}
	\label{fig:MW_mass_assembly}
\end{figure}

\reffig{fig:MW_mass_assembly} shows the mass growth history of the DM
halo and of the central galaxy in the five MW analogues studied in
\refsec{sect:MW-like_examples}. The central galaxies have assembled
most of their mass by $z=1$ (except MW1 and MW3 which have a slightly
later formation time), after which they experience only a modest
growth in stellar mass.

It is interesting to contrast \reffig{fig:MW_mass_assembly} with the
changes in galaxy and halo orientation shown in \reffig{fig:MW_time}.
The orientation of the central galaxies can vary considerably during
the phase of rapid stellar growth ($z>1$); however, at later times,
when the mass growth is slower, the orientation remains nearly
constant. In contrast, the orientation of the DM halos can vary
significantly even at $z<1$ when their growth rate has slowed down.

\section{The orbits of LMC analogues}
\label{sec:app_LMC_orbits}
\begin{figure}
    \centering
	\plotone{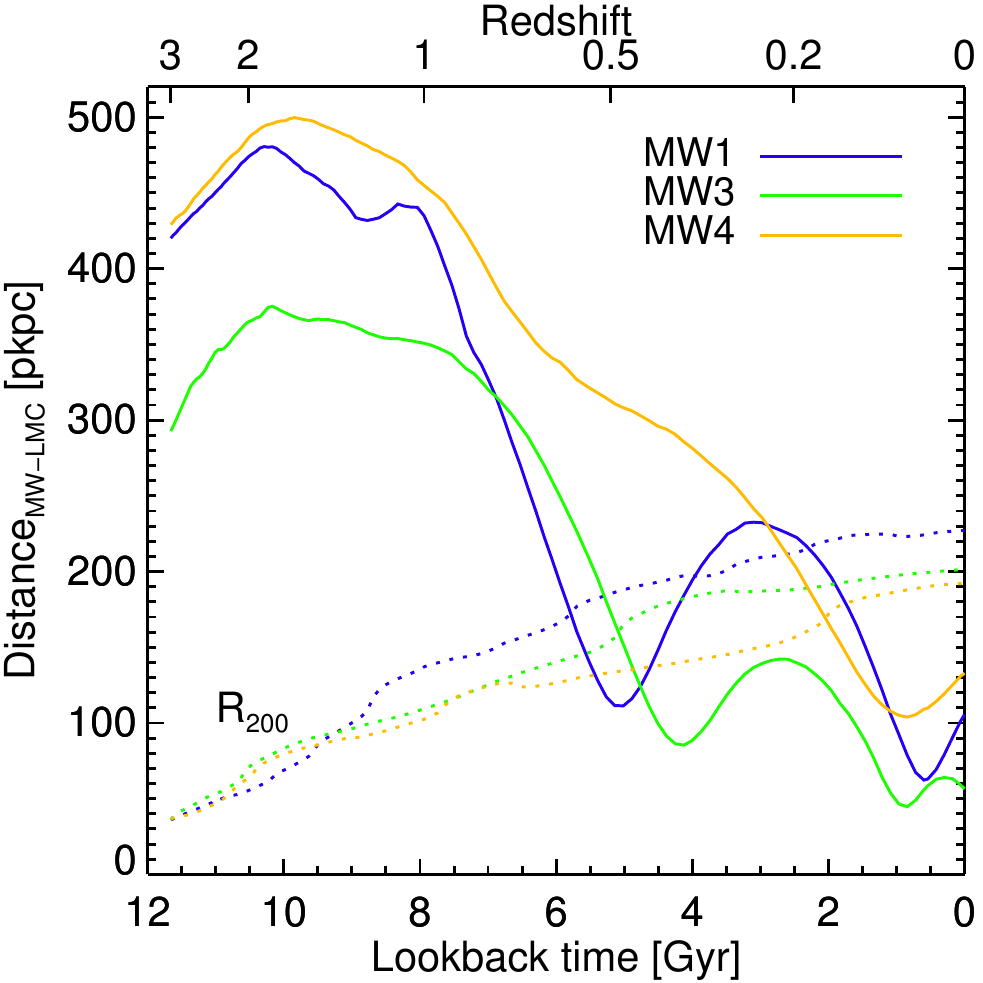}
	\caption{The distance between the LMC-mass dwarf and the
          progenitor of the $z=0$ MW-mass host halo. The solid lines
          correspond to each of the three MW analogues that contain an
          LMC-mass satellite. The dotted lines show the radius,
          $R_{\rm 200}$, of each of the three host haloes.  }
	\label{fig:LMC_orbits}
\end{figure}
\reffig{fig:LMC_orbits} shows the orbits of the three LMC-mass
satellites we found in our sample of analogues of the MW bright
satellite population. In two of the systems, MW1 and MW3, the LMC-mass
satellite has just passed its second pericentre, while in MW4 the
massive satellite has just passed its first pericentre.

It is instructive to compare the accretion times of the LMC analogues,
that is the time when they first crossed the host halo radius, with
the time when the host experienced its last large change in orientation
(see \reffig{fig:MW_time}). The three LMC-mass satellites
were accreted 6, 5.5, and 3 Gyrs ago, while their host haloes retained
a roughly constant orientation (i.e. $\cos\theta>0.8$ in the bottom
panel of \reffig{fig:MW_time}) from 8, 5, and 4 Gyrs ago,
respectively. Thus, the accretion of the LMC-mass satellite occurred 
around the same time as the last major reorientation of their MW-mass host halo.

\label{lastpage}
\end{document}